\documentclass[%
 reprint,
nofootinbib,
 amsmath,amssymb,
 aps,
]{revtex4-2}

\usepackage{amsmath}
\usepackage{mathtools}
\usepackage{graphicx}
\usepackage{dcolumn}
\usepackage{bm}
\usepackage[normalem]{ulem}
\usepackage[com pat=1.1.0]{tikz-feynman}
\usepackage{tikz}
\usepackage{comment}
\usetikzlibrary{arrows}
\usepackage[colorlinks=true, allcolors=blue]{hyperref}
\usepackage{aas_macros}
\usetikzlibrary{external}
\begin{document}

\preprint{APS/123-QED}

\title{Effects of energy-dependent scatterings on fast neutrino flavor conversions}

\author{Chinami Kato}
 \affiliation{Faculty of Science and Technology, Tokyo University of Science, 2641 Yamazaki, Noda-shi, Chiba Prefecture 278-8510, Japan}
 \email{ckato@rs.tus.ac.jp}
\author{Hiroki Nagakura}%
\affiliation{%
 National Astronomical Observatory of Japan, 2-21-1 Osawa, Mitaka, Tokyo 181-8588, Japan}%

\date{\today}

\begin{abstract}
Neutrino self-interactions in a dense neutrino gas can induce collective neutrino flavor conversions.
Fast neutrino flavor conversions (FFCs), one of the collective neutrino conversion modes, potentially change the dynamics and observables in core-collapse supernovae and binary neutron star mergers.
In cases without neutrino-matter interactions (or collisions), FFCs are
essentially energy-independent, and therefore the single energy treatment has been used in previous studies.
However, neutrino-matter collisions in general depend on neutrino energy, suggesting that energy-dependent features may emerge in FFCs with collisions.
In this paper, we perform dynamical simulations of FFCs with iso-energetic scatterings (emulating nucleon scatterings) under multi-energy treatment.
We find that cancellation between in- and out-scatterings happens in high energy region, which effectively reduces the number of collisions and then affects the FFC dynamics.
In fact, the lifetime of FFCs is extended compared to the single-energy case, leading to large flavor conversions.
Our result suggests that the multi-energy treatment is mandatory to gauge the sensitivity of FFCs to collisions.
We also provide a useful quantity to measure the importance of multi-energy effects of collisions on FFCs.
\end{abstract}

\maketitle

\section{Introduction}
In dense neutrino environments such as core-collapse supernovae (SNe), binary neutron star mergers (BNSMs), and early Universe, refractive effects by neutrino self-interactions can induce collective neutrino flavor conversions
\cite{samuel1993,sigl1993,sigl1995,sawyer2005}.
Under certain conditions, the growth rate of flavor conversion becomes proportional to the number density of neutrinos, which corresponds to a fast mode or fast flavor conversion (FFC).
FFC is triggered when electron neutrino lepton number crossings (ELN crossings) appear in angular distributions of neutrinos \cite{morinaga2021}.
Recent studies have indicated that ELN crossings are ubiquitous in SNe and BNSMs \cite{dasgupta2017,tamborra2017,wu2017,dasgupta2018,abbar2019b,nagakura2019,milad2019,glas2020,abbar2020,abbar2020b,morinaga2020,milad2020,capozzi2020,abbar2021,capozzi2021,nagakura2021d,harada2022, sherwood2022, akaho2022}.
Since neutrino flavor conversions, in general, have a potential to affect the fluid dynamics, nucleosynthesis, and neutrino signals
\cite{wu2017,george2020,Li2021,Just2022}, FFCs have garnered significant attention recently.

In SNe and BNSMs, neutrinos also undergo matter interactions (or collisions) during propagation through matter.
The description of neutrino dynamics involving neutrino transport, flavor conversion, and neutrino-matter collisions requires quantum kinetic treatment.
SN- and BNSM simulations with quantum kinetic neutrino transport are, however, a formidable challenge for computational astrophysics.
In fact, their global simulations with resolving small-scale variations of FFCs (see, e.g., \cite{abbar2019,martin2020,johns2020b,bhattacharyya2020,bhattacharayya2021,sherwood2021,sherwood2021b,wu2021}) are intractable at present.
Currently feasible approach is either local simulations \cite{abbar2019,sherwood2019,martin2020,bhattacharyya2020,shalgar2021,sherwood2021,sherwood2021b,martin2021,zaizen2021,bhttacharyya2022,Abbar2022} or global simulations with some simplifications and assumptions \cite{Li2021,nagakura2022,shalgar2022,shalgar2022b}.

Recently effects of collisions on FFCs have been also studied numerically \cite{shalgar2021b,martin2020,kato2021,sasaki2021,sigl2022,hansen2022}.
Some studies have found enhancement of FFCs by neutral-current-like matter collisions \cite{shalgar2021b,kato2021,sasaki2021,hansen2022}, whereas others have reported damping effects \cite{martin2020,sigl2022}.
Although these results are apparently in conflict with each other, which may be due to different angular distributions of neutrinos \cite{hansen2022} or self-consistency issues faced by homogeneous models \cite{lucas2022}, they are in agreement with a suggestion that collisions have non-negligible roles on FFCs.
We also note that a new type of instability of flavor conversion, collisional flavor instability, may occur due to disparity of matter interactions between neutrinos and antineutrinos \cite{johns2021}.

FFCs are purely driven by neutrino self-interactions, suggesting that energy dependent features induced by vacuum potential are smeared out\footnote{It should be noted that the vacuum potential plays a role as a seed perturbation to trigger FFCs.}.
Hence, the energy-integrated QKEs or monochromatic assumption has been commonly used in the previous FFC studies.
However, the approximation becomes inappropriate if neutrino-matter interactions are involved. Cross sections of weak current depend on energy, and the energy spectrum of neutrinos is not monochromatic in SNe/BNSMs, indicating that energy-dependent flavor conversions may be induced by collisions.
In fact, recent studies showed that vigor of FFCs hinges on the reaction rate \cite{hansen2022}, and therefore it is of great interest to investigate FFCs with energy-dependent collisions in non-monochromatic neutrino energy spectrum.

In this paper, we perform
dynamical simulations of FFCs in homogeneous background with incorporating energy-dependent collisions (iso-energetic scatterings).
The goal of this paper is to 
understand the essence of multi-energy effects of iso-scatterings on FFCs. In so doing, a simple setup is adopted in our simulations. We find some qualitative differences of FFCs between single- and multi-energy treatments in the non-linear phase, which is scrutinized in this paper.

The paper is organized as follows. We start with describing basic equations and our numerical solver with a Monte Carlo method in Section~\ref{sec:setup}.
Before entering into results of multi-energy FFC simulations, we revisit the case of those with energy-independent collisions in Section~\ref{sub:scat_effect}, which provides some clues to interpret multi-energy effects of collisions on FFCs.
We then discuss FFCs with energy-dependent collisions in two cases: flat neutrino energy spectrum and non-flat spectrum in Sections~\ref{sub:multi_effect} and \ref{sub:asym_effect}, respectively.
In Section~\ref{sub:threemeshes}, we present an energy-resolution study to assess our proposed diagnostics for quantifying multi-energy effects.
Finally, we conclude and discuss our study in Section~\ref{sec:summary}.

\section{Numerical setup} \label{sec:setup}
\subsection{Basic equations}
Time evolution of neutrinos in phase space follows the quantum kinetic equations (QKEs),
\begin{eqnarray}
i\left( \frac{\partial}{\partial t} + \vec{v}\cdot \nabla \right) \rho(\vec{x},\vec{p},t) &=& \left[\mathcal{H}, \rho(\vec{x},\vec{p},t)\right] 
+ i \mathcal{C[\rho]}, \label{eq:rho_orig}\\
i\left( \frac{\partial}{\partial t} + \vec{v}\cdot \nabla \right) \bar{\rho}(\vec{x},\vec{p},t) &=& \left[\bar{\mathcal{H}}, \bar{\rho}(\vec{x},\vec{p},t)\right]
+ i \mathcal{\bar{C}[\bar{\rho}]}. \label{eq:rhob_orig}
\end{eqnarray}
In this expression, $\rho$ $(\bar{\rho})$ describes the density matrix for neutrinos (antineutrinos). 
The Hamiltonian potentials $\mathcal{H}$ and $\bar{\mathcal{H}}$ include contributions from the neutrino mass difference, the coherent forward scattering with surrounding matters and other neutrinos as detailed in \cite{sigl1993,volpe2015}.
$\mathcal{C}$ and $\bar{\mathcal{C}}$ represent the collision terms.

We assume a two flavor system comprised of electron-type neutrinos ($\nu_e$) and heavy-leptonic ones ($\nu_x$).
In this approximation, the density matrix has four independent components, i.e., $\rho_{ee}$, $\rho_{xx}$, ${\rm Re}\rho_{ex}$ and ${\rm Im}\rho_{ex}$.
We set the mass difference of neutrinos to be 0, and the matter term is also neglected, just for simplicity.
It is also assumed that neutrino distributions are spatial homogeneous and axial asymmetry in momentum space.
For collision terms, we consider iso-energetic scatterings.
The reaction rate depends on neutrino energy, but we assume that it is independent on propagation direction (i.e., isotropic scattering).
The rates are also assumed to be the same among all the components of density matrices.
Under these assumptions, QKEs can be rewritten as
\begin{eqnarray}
&& i \frac{\partial \rho_a(E_\nu,\cos{\theta_\nu},t)} {\partial t} \nonumber \\
&& = \left[\mathcal{H}_{\nu\nu}, \rho_a(E_\nu,\cos{\theta_\nu},t)\right] \nonumber \\
&& + i  \int_{-1}^1 d\cos{\theta_\nu^\prime}R(E_\nu) \rho_a(E_\nu,\cos{\theta_\nu^\prime,t}) \nonumber \\
&& - i  \int_{-1}^1 d\cos{\theta_\nu^\prime}R(E_\nu) \rho_a(E_\nu,\cos{\theta_\nu},t), \label{eq:rho}\\
&& i \frac{\partial \bar{\rho}_a(E_\nu,\cos{\theta_\nu},t)} {\partial t} \nonumber \\
&& = \left[\bar{\mathcal{H}}_{\nu\nu}, \bar{\rho}_a(E_\nu,\cos{\theta_\nu},t)\right] \nonumber \\
&& + i  \int_{-1}^1 d\cos{\theta_\nu^\prime}R(E_\nu) \bar{\rho}_a(E_\nu,\cos{\theta_\nu^\prime,t}) \nonumber \\
&& - i  \int_{-1}^1 d\cos{\theta_\nu^\prime}R(E_\nu) \bar{\rho}_a(E_\nu,\cos{\theta_\nu},t).
\label{eq:rhob}
\end{eqnarray}
In this expression, $\rho_a = \int \rho d\phi_\nu$ denotes the azimuthal-angle integrated density matrix. The Hamiltonians by neutrino self-interactions can be written as
\begin{eqnarray}
\mathcal{H}_{\nu\nu} &=& \sqrt{2}G_F \int\int \frac{E_\nu^{\prime 2}dE_\nu^\prime d\cos{\theta_\nu^\prime}}{(2\pi)^2} 
\nonumber \\
&\times& (1-\cos{\theta_\nu}\cos{\theta_\nu^\prime}) \nonumber \\ 
&\times& (\rho_a(E_\nu,\cos{\theta_\nu^\prime},t)-\bar{\rho}_a^{\ast}(E_\nu,\cos{\theta_\nu},t)), \\
\bar{\mathcal{H}}_{\nu\nu} &=& \sqrt{2}G_F \int \int \frac{E_\nu^{2\prime}dE_\nu^\prime d\cos{\theta_\nu^\prime}}{(2\pi)^2} \nonumber \\
&\times& (1-\cos{\theta_\nu}\cos{\theta_\nu^\prime}) \nonumber \\
&\times& (\rho^{\ast}_a(E_\nu,\cos{\theta_\nu^\prime},t)-\bar{\rho}_a(E_\nu,\cos{\theta_\nu},t)).
\end{eqnarray}
Regarding the reaction kernel $R(E_\nu)$, the energy dependence emulates nucleon scatterings, i.e., $R(E_\nu)=R_0(E_\nu/20{\rm MeV})^2$.
In this study, we adopt
$\mu \equiv \sqrt{2}G_Fn_{\nu} = 10\ {\rm cm^{-1}}$, where $n_{\nu}$ denotes the number density of $\nu_e$. This corresponds to $n_{\nu} \sim 1.56\times10^{33}\ {\rm cm^{-3}}$.

We refer \cite{shalgar2021b} to set initial angular distributions of neutrinos.
$\nu_e$ have isotropic angular distributions, whereas antineutrinos ($\bar{\nu}_e$) have anisotropic ones, 
\begin{eqnarray}
g_{ee,0}(\cos{\theta_\nu}) &=& 0.5n_{\nu}, \nonumber \\
\bar{g}_{ee,0}(\cos{\theta_\nu}) &=& \left[0.47 + 0.05\exp{\left(-\left(\cos{\theta_\nu}-1\right)^2\right)}\right]n_\nu, \label{eq:initial_angle}\ \ \ \ 
\end{eqnarray}
where $g$ is defined as
$g \equiv \int \rho_a E_\nu^2 dE_\nu/(2\pi)^3$.
Those for $\nu_x$ and their antipartners ($\bar{\nu}_x$) are set be 0 initially. At the beginning of simulations, we put a small perturbation on off-diagonal components of $\rho_a$ by ${\rm Im}g_{ex} = -{\rm Re}g_{ex}=10^{-6}g_{ee}$, to trigger FFCs.
Note that the number density of anti-neutrinos $\bar{n}_{\nu}$ is $1.53\times10^{33}\ {\rm cm^{-3}}$, which is $\sim 2 \%$ smaller than $n_\nu$.

\subsection{QKE-MC solver}
We solve QKEs (Eqs.~\ref{eq:rho}~and~\ref{eq:rhob}) using Monte-Carlo technique \cite{kato2021}.
Our code is an extended version from our classical MC neutrino transport solver \cite{kato2020}.
This QKE-MC solver has a capability of handling neutrino transport, matter collisions and neutrino flavor conversions self-consistently.
Below we describe the essence of our code.

In classical MC methods, each MC particle represents a bundle of neutrinos in a specific flavor state, and the assembly of these particles describes a neutrino distribution function in phase space.
On the other hand, if flavor conversions happen, flavor states are no longer constant during propagation.
Hence, we add a flavor degree of freedom to each MC particle.
More specifically, we define a new matrix, particle-density-matrix (PDM), which is assigned to each MC particle; the assembly of the PDM represents the density matrix of neutrinos.

In our QKE-MC solver, we handle collisions and flavor conversions separately, which is essentially along with an operating-splitting method.
Firstly, we determine what particles and their flavor states undergo scatterings \footnote{We assign a scattering distance to each element of the PDM.}.
We create a new MC particle at each scattering point, and its flight direction is determined probabilistically with a scattering kernel.
In the meantime, we compute the self-interaction potential of neutrinos, and then evolve the PDM on each MC particle with fourth-order Runge-Kutta method.
It should be mentioned that feedback from collisions is not included here. After updating the PDM, we copy the scattering-experienced PDM from the original MC particle to newly created ones.

Statistical noise, which is inevitably generated by probabilistic computations of collisions in MC methods, is a major concern.
This may compromise the accuracy of our simulations.
To mitigate the problem without substantially increasing computational costs, we have developed a special technique.
Firstly, we discretize the neutrino phase space similar to mesh-based methods. The flight direction of each MC particle is corrected at each timestep to match with the central value of the mesh-point where the particle resides.
Although this prescription smears out a small scale structure in neutrino momentum space, it can be controlled by changing the mesh resolution.
Secondly, the effective mean free path (EMFP) method is adopted.
In essence, we reduce the rate of change in the PDM ($a$ ($0\leq a \leq 1$)) at each scattering event, meanwhile the scattering length is also reduced by a factor of $a$. This prescription increases the number of scattering of MC particles (hence, the statistics is improved) with sustaining the physical impact of collisions. It should be mentioned that the mean free time of collisions is much longer than the FFC timescale, implying that the number of scatterings during a single timestep is small. As a result, the numerical cost of computing collisions is subdominant compared to other operations; hence we can maintain the computational cost even if the number of scatterings is increased by the EMFP prescription.
We refer readers to \cite{kato2021} for more details of these techniques and a suite of code tests.

In this study, we deploy 128 bins in angular direction of neutrinos and adopt 50,000 MC samples in each angular bin.
The control parameter of EMFP method is set to be $a=10^{-3}$, which shows a good performance to reduce the statistical noise without increasing numerical cost. 
In multi-energy cases, we run several simulations with two energy meshes (see Sections~\ref{sub:multi_effect}~and~\ref{sub:asym_effect}) and with three ones (Section~\ref{sub:threemeshes}), which are useful to highlight the essential multi-energy effects of collisions on FFCs.

\begin{figure}
\hspace*{-1cm}
\includegraphics[width=9cm]{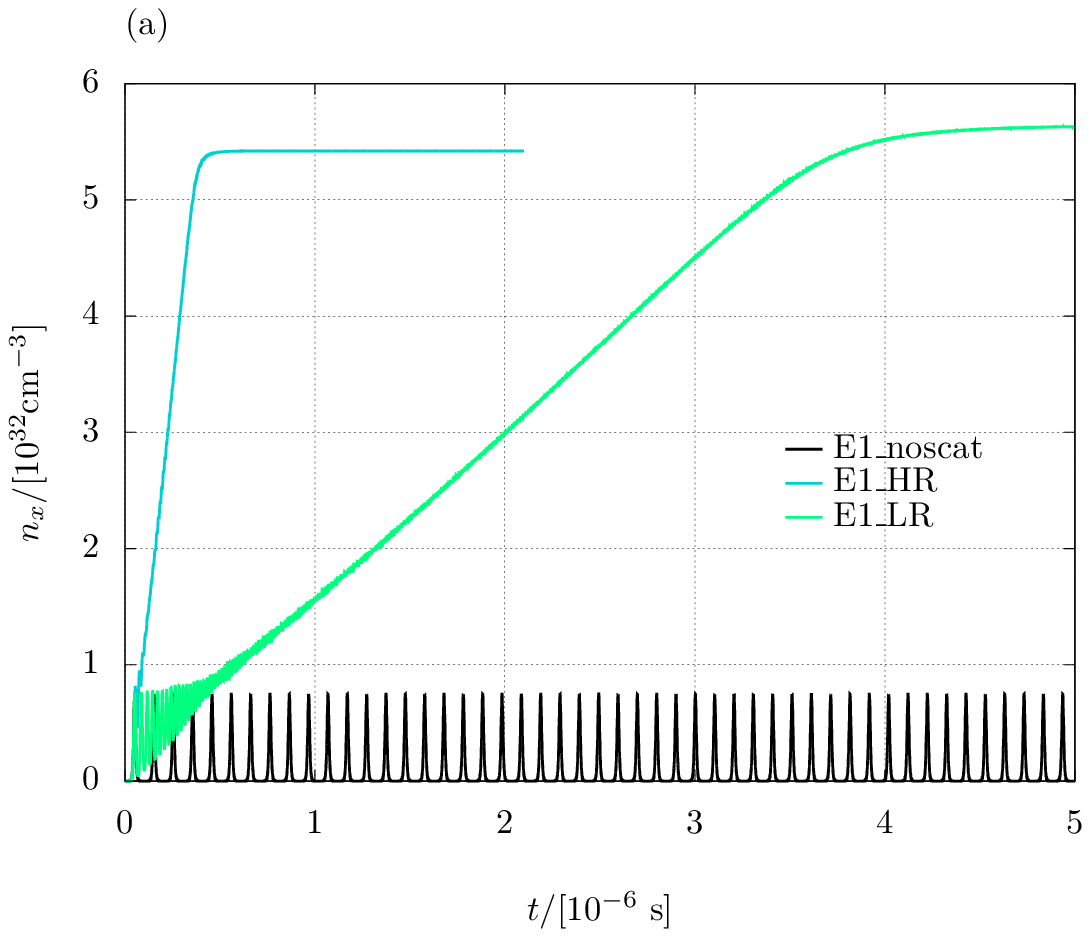}
\hspace*{-1cm}
\includegraphics[width=9cm]{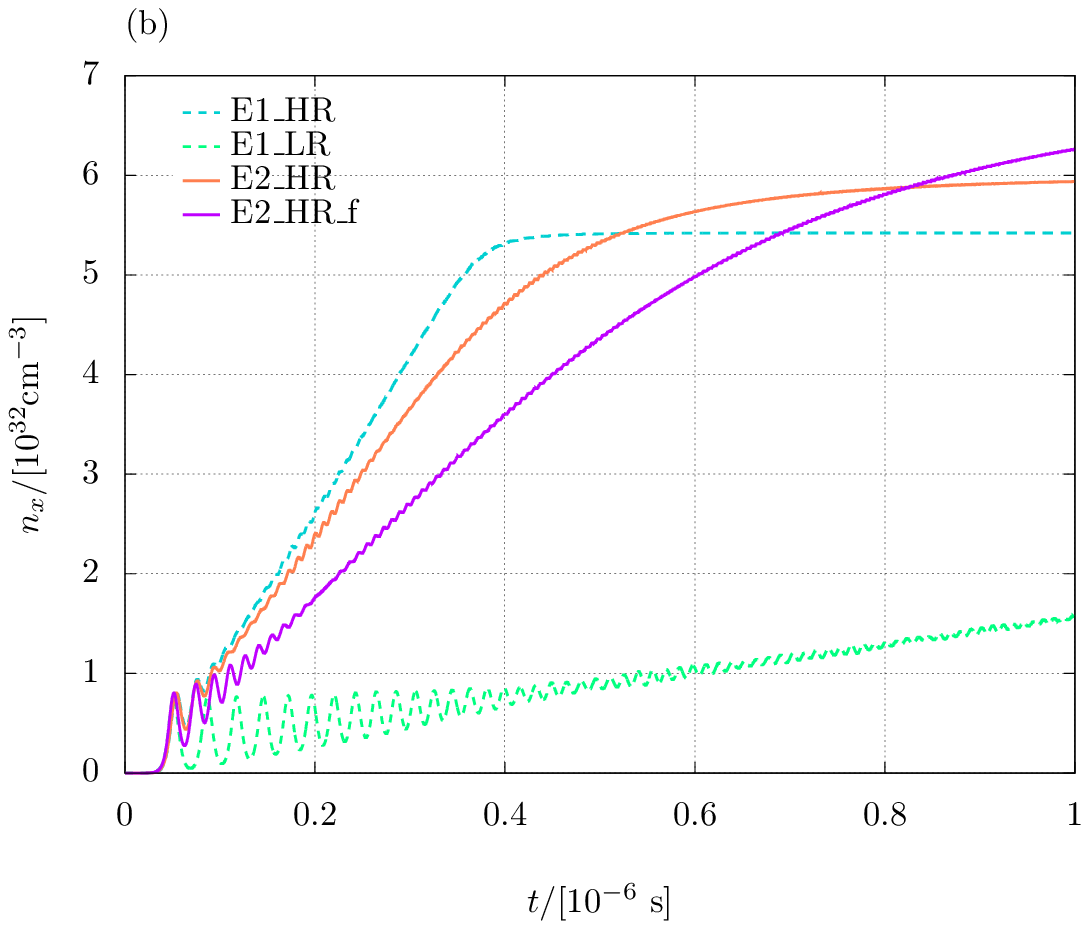}
\hspace*{-1cm}
\includegraphics[width=9cm]{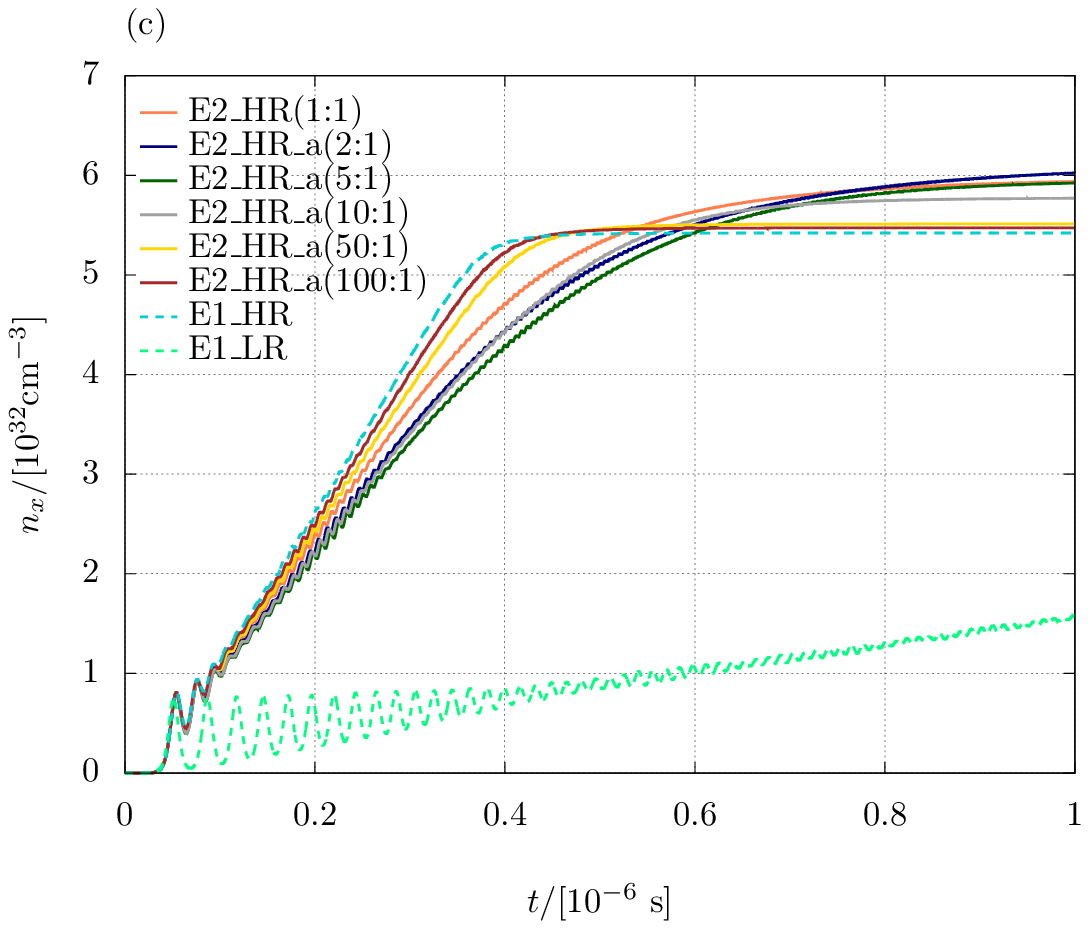}
\caption{\label{fig:rhoxx_scateffect} Time evolution of $n_x$ for (a) E1, (b) E2$\_$HR and (c) E2$\_$HR$\_$a models.
Black line in panel(a) describes the result without collisions. Other colors distinguish the models with collisions. In panels(b) and (c), the solid and dashed-lines represent E2 and E1 models, respectively. }
\end{figure}

\section{results}
In this study, we have 10 models in total with different numerical setups. In Section~\ref{sub:scat_effect}, we first show results of FFC simulations with energy-independent collisions, in which we adopt a single-energy mesh.
We then present results of multi-energy FFC simulations with energy-dependent collisions in Sections~\ref{sub:multi_effect}-\ref{sub:threemeshes}.

\begin{figure*}
\begin{flushleft}
\hspace*{-1.5cm}
\includegraphics[width=20cm]{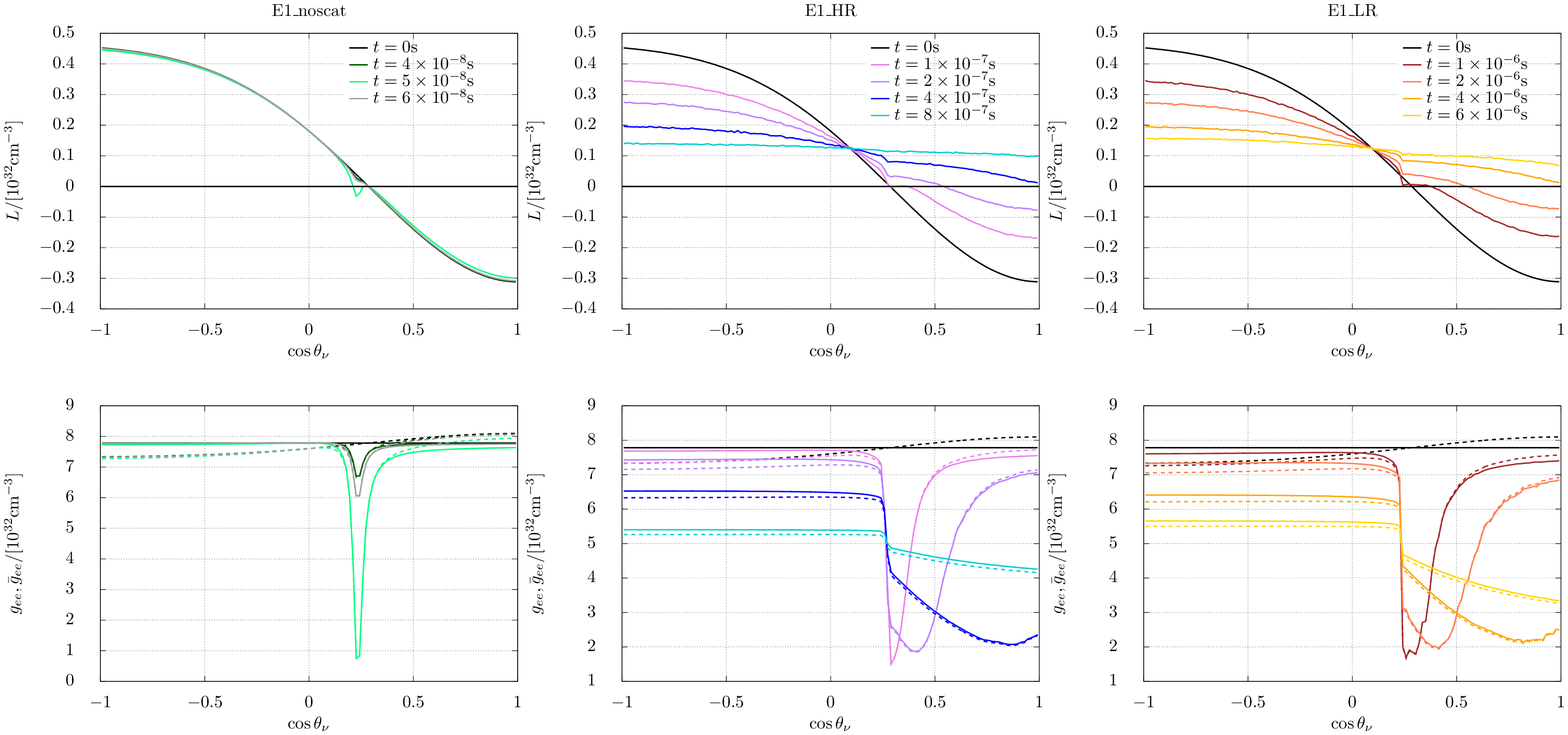}
\caption{\label{fig:eln} Angular distributions of $L$ (top) and $g_{ee}$ (bottom) for E1 models. From the left panels, we show the results of E1$\_$noscat, E1$\_$HR and E1$\_$LR models, respectively. Different colors denote different timesteps.}
\end{flushleft}
\end{figure*}

\subsection{Energy-independent collisions} \label{sub:scat_effect}
Here we revisit effects of energy-independent collisions on FFCs.
We consider three cases with different reaction rates: $R_0 = 0$ (noscat), $1.25\times10^{-5}$ (LR), $1.25\times10^{-4}\ {\rm cm^{-1}}$ (HR), assuming $R(E_\nu)=R_0(E_\nu/20{\rm MeV})^2$ with $E_\nu=20$ MeV.
These models are referred to E1 models hereafter.

Figure~\ref{fig:rhoxx_scateffect}(a) portrays time evolution of $\nu_x$'s number density, or $n_x\equiv\int d\cos{\theta_\nu} g_{xx}$ for E1 models.
Since $g_{xx}$ is initially zero (see Section~\ref{sec:setup}), the appearance of $\nu_x$ exhibits occurrence of flavor conversions.
We also note that the trace of density matrix is not affected by flavor conversions, and iso-energetic scattering does not change the total number of neutrinos, implying that $n_e+n_x$ is a conserved quantity in our models.
Black line describes the results without collisions (noscat), in which the time evolution of $n_x$ has a periodic feature with a constant amplitude.
This is consistent with previous studies \cite{shalgar2021b}, and this feature may be understood through a pendulum model \cite{Ian2022,johns2020}.
In cases with collisions, on the other hand, the periodic feature disappears. Although the amplitude of oscillations is smaller than that in the case without collisions, $n_x$ increases with time, and eventually being saturated at a certain point (see cyan and green lines in Figure~\ref{fig:rhoxx_scateffect}(a)).
In these models, the saturation densities of $\nu_x$ become higher than the peak of $\nu_x$ in the case without collisions, indicating that collisions enhance flavor conversions.

The top panels in Figure~\ref{fig:eln} show
angular distributions of ELN-XLN, or $L(\cos{\theta_\nu}) \equiv g_{ee}(\cos{\theta_\nu})-g_{xx}(\cos{\theta_\nu})-\bar{g}_{ee}(\cos{\theta_\nu})+\bar{g}_{xx}(\cos{\theta_\nu})$ at some selected times. Here XLN denotes a heavy-leptonic neutrino lepton number.
As discussed in \cite{nagakura2022}, ELN-XLN angular distributions characterize the growth of FFCs\footnote{There is a caveat, however. In homogeneous models, the appearance of ELN or ELN-XLN crossings does not guarantee the occurrence of FFC. This is due to the fact that the instability would occur in inhomogeneous modes, while flavor conversion may be stable in homogeneous ones. Nevertheless, ELN-XLN angular distributions are still informative to understand qualitative trends of FFCs in homogeneous models. In fact, the disappearance of ELN-XLN angular crossings guarantees that FFCs do not occur in homogeneous models.}.
In the case without collisions (left-top panel), the angular distribution of $L$ does not evolve much with time except in the vicinity of ELN-XLN crossing.
As a result, the ELN-XLN crossing does not disappear throughout the evolution.
In cases with collisions, on the other hand, the angular point where the ELN-XLN crossing occurs changes with time, and eventually the crossing disappears ($L$
becomes positive in all directions).
Around the time when the ELN-XLN crossing disappears ($t\sim4\times10^{-7}$, $4\times10^{-6}$ s for E1$\_$HR and E1$\_$LR models, respectively), $n_x$ is saturated, indicating that FFCs subside.
Since angular distributions of neutrinos are still anisotropic at the saturation time, they continue to be isotropized by collisions. 
The isotropization does not generate new ELN-XLN crossings, and therefore FFCs are no longer revived.

One of the important roles of collisions on FFC is to broaden the angular region where FFC occurs.
As shown in left bottom panel of Figure~\ref{fig:eln},
$g_{ee}$ $(\bar{g}_{ee})$ in E1$\_$noscat model decreases due to FFCs only around the region where $L$ is zero ($\cos{\theta_\nu}\sim0.2$).
For cases with collisions, on the other hand, FFCs occur in the wider angular range (see middle- and right bottom panels of Figure~\ref{fig:eln}). This is consistent with our observation that the ELN-XLN crossing point moves to forward angular directions due to isotropization by collisions.

\begin{figure}
\hspace*{-1.0cm}
\includegraphics[width=8.5cm]{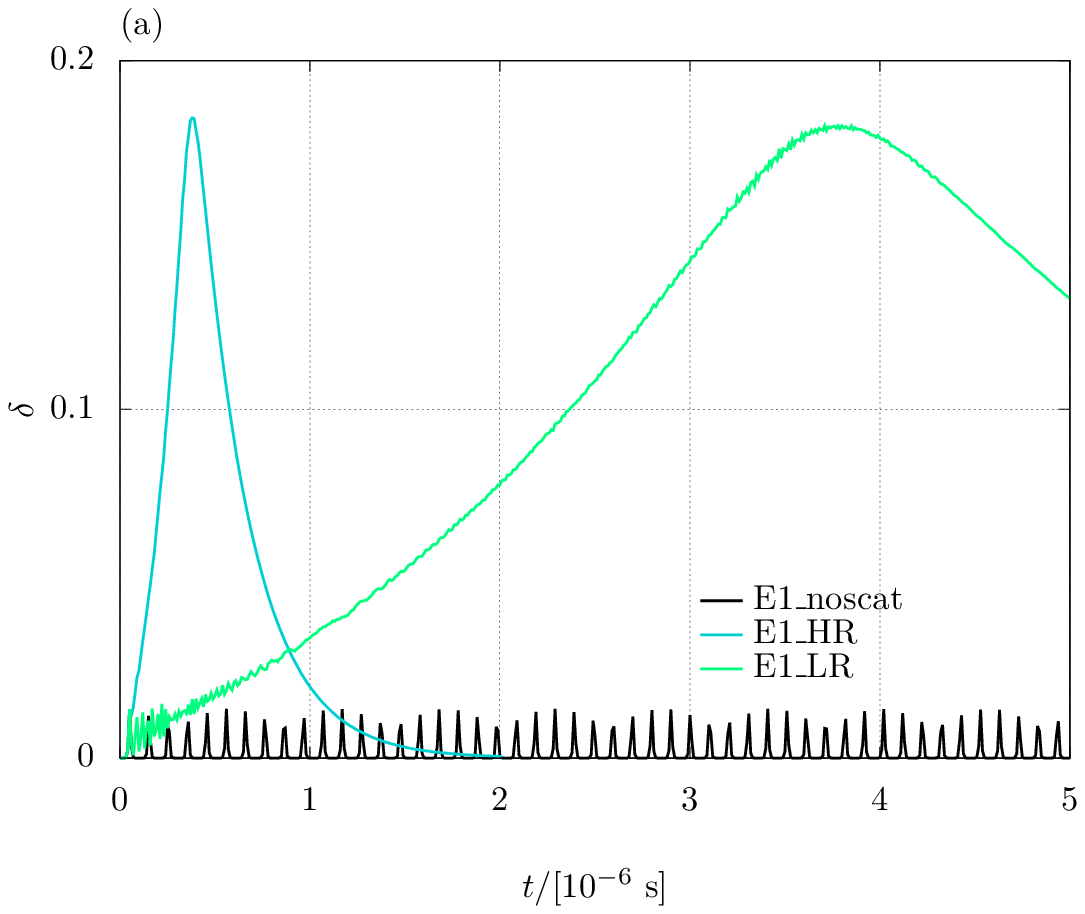}
\hspace*{-1.0cm}
\includegraphics[width=8.5cm]{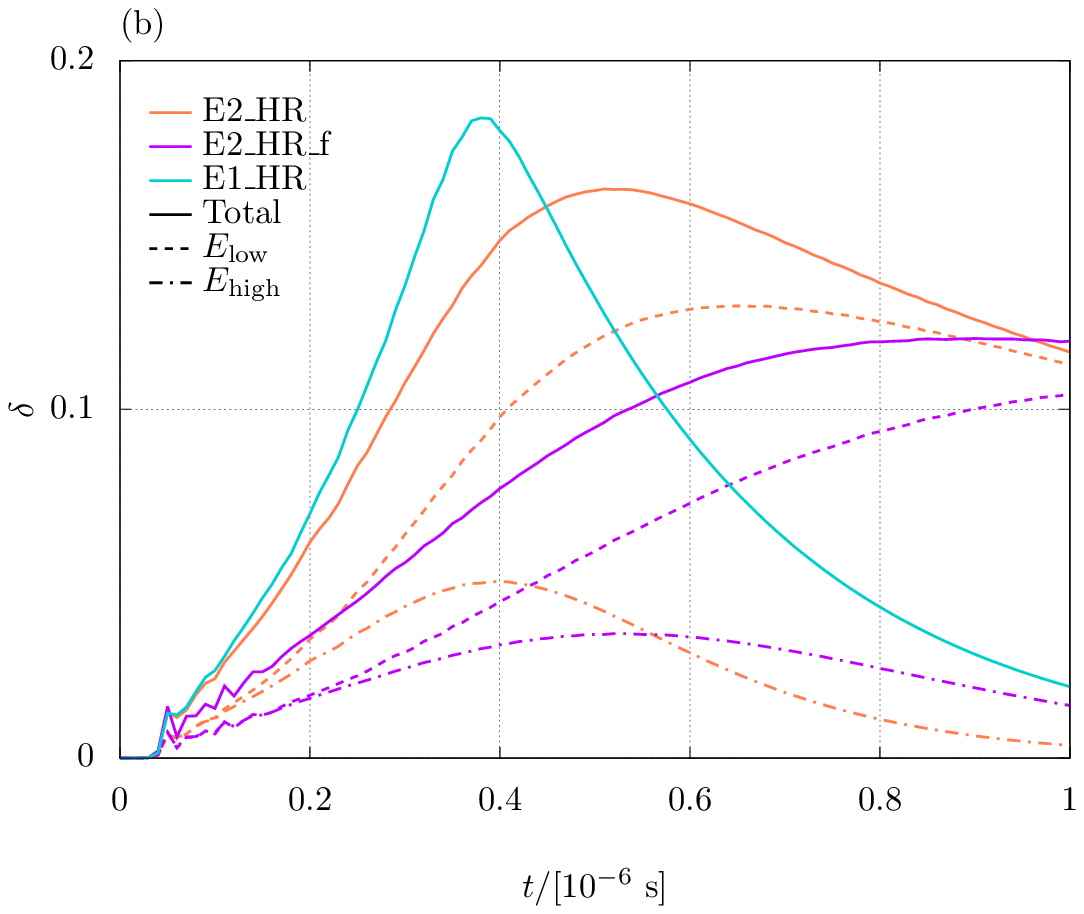}
\hspace*{-1.0cm}
\includegraphics[width=8.5cm]{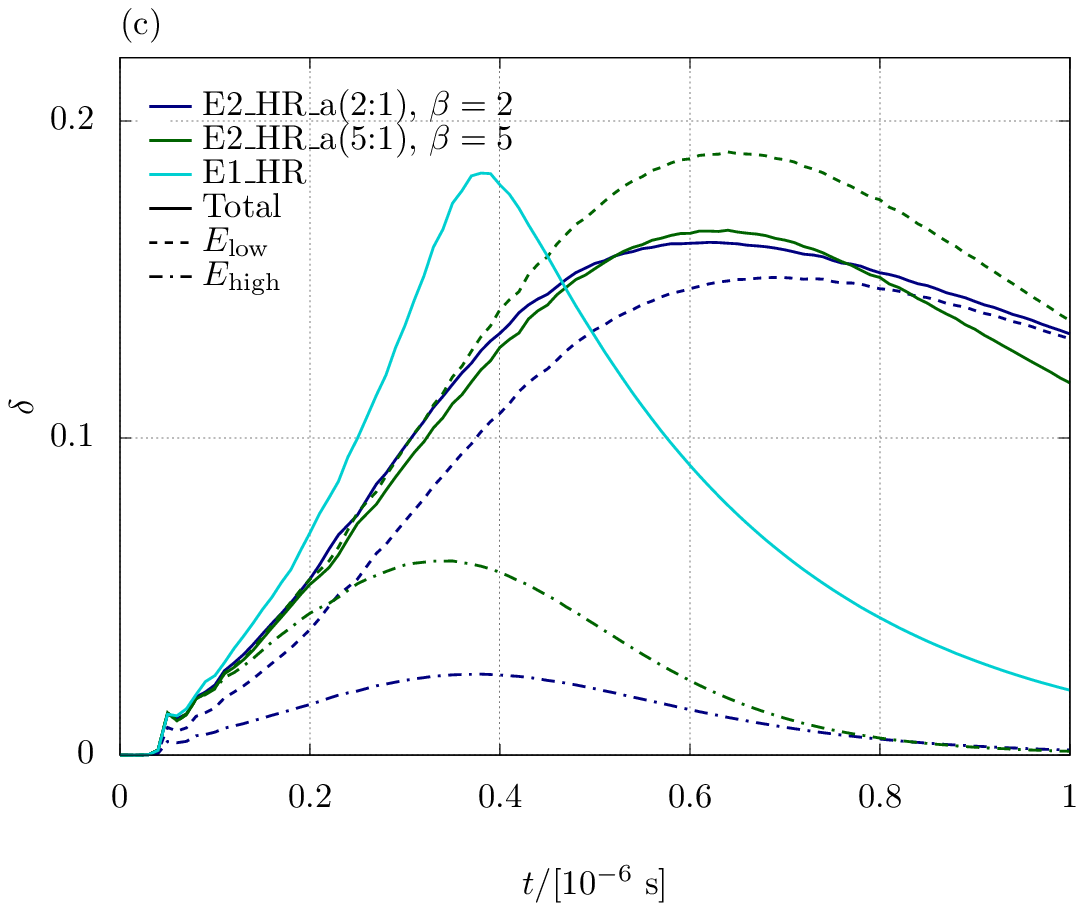}
\caption{\label{fig:deg_asym} Time evolution of $\delta$ for (a) E1, (b) E2$\_$HR and (c) two selected E2$\_$HR$\_$a models. 
Colors distinguish the models. Solid lines describe $\delta$ with energy-integrated angular distributions in eq.\ref{eq:anisotropy} (Total), while the dashed and dash-dotted lines denote those of low- ($E_{\rm low}$) and high-energy neutrinos ($E_{\rm high}$), respectively. For the low- (high-) energy case, we
replace $g_{ee}$ to $g_{ee,{\rm low}}$ $(g_{ee,{\rm high}})$ in the numerator of Eq.\ref{eq:anisotropy}.}
\end{figure}

To measure the anisotropy of $\nu_e$, we define $\delta$ as
\begin{eqnarray}
 \delta = \left| \frac{\int d\cos{\theta_\nu} \cos{\theta_\nu}g_
 {ee}(\cos{\theta_\nu})}{\int d\cos{\theta_\nu} g_{ee}\left(\cos{\theta_\nu}\right)}\right|,
 \label{eq:anisotropy}
\end{eqnarray}
and its time evolution is displayed in Figure~\ref{fig:deg_asym}(a).
In the case without collisions (black line), $\delta$ has a periodic feature, which is the same trend as $n_x$.
Green and cyan lines depict the results for E1$\_$LR and E1$\_$HR models, respectively.
In these models, $\delta$ increases due to FFCs until the disappearance of the ELN-XLN crossing. After FFCs are terminated, $\delta$ monotonically decreases by collisions.

As shown in Figure~\ref{fig:rhoxx_scateffect}(a), we find that 
the vigor of FFCs is stronger with increasing reaction rate, meanwhile the saturation amplitude and its time become smaller and shorter, respectively; those for E1$\_$HR model (E1$\_$LR model) are $n_x\sim 5.5\times10^{32}\ {\rm cm^{-3}}$ and $t\sim4\times10^{-7}$ s ($n_x\sim5.6\times10^{32} \ {\rm cm^{-3}}$ and $t\sim4\times10^{-6}$ s).
This exhibits that the saturation amplitude of FFCs is determined through the competition between the growth rate and the duration of FFCs. This trend needs to be kept in our mind in the following discussions. It should also be mentioned that our results presented in this section is consistent with previous studies \cite{shalgar2021b,sasaki2021,hansen2022}. In the following subsections, we turn our attention to multi-energy effects of collisions on FFCs, which is the subject of this paper.

\begin{figure*}
\begin{flushleft}
\hspace*{-1.5cm}
\includegraphics[width=20cm]{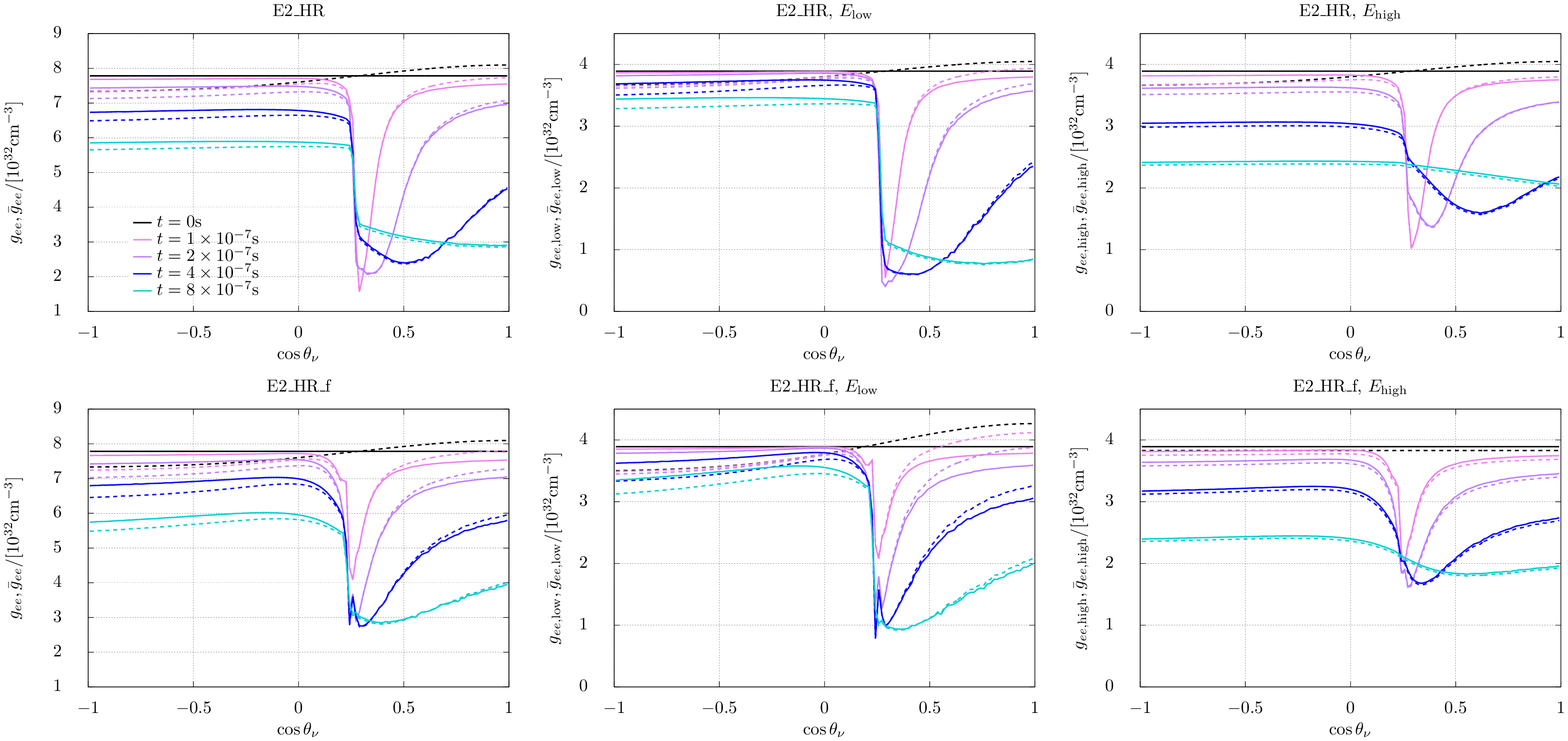}
\caption{\label{fig:angle_tot_E2} Angular distributions for E2$\_$HR (top), E2$\_$HR$\_$f models (bottom). The energy-integrated distributions are shown in the left panels, and those for low- and high
neutrino energy are plotted in the middle and right panels, respectively. Different colors denote the different timesteps, which are the same as those selected in Figure~\ref{fig:eln} (for E1$\_$HR model). Solid and dashed lines describe $\nu_e$ and $\bar{\nu}_e$, respectively.}
\end{flushleft}
\end{figure*}

\subsection{Energy-dependent collisions: flat neutrino-energy spectrum} \label{sub:multi_effect}

Here, we discuss multi-energy effects of collisions on FFCs with comparing to the case of energy-independent collisions presented in Section~\ref{sub:scat_effect}.
For the sake of simplicity, neutrino energies are assumed to be 
$E_{\rm low}=10$~MeV 
or
$E_{\rm high}=30$~MeV.
We refer to the models as E2.
In this section, we consider the case with flat neutrino-energy spectrum, i.e., $n_{\rm low}=n_{\rm high}=n_\nu/2$
where $n_{\rm low}$ and $n_{\rm high}$ denote the numbers of neutrinos with the energy of $E_{\rm low}$ and $E_{\rm high}$, respectively.
We note that the case of non-flat energy spectrum will be discussed in the next subsection.

The angular distributions of neutrinos are set to be identical to those used in previous subsection (see Eqs.~\ref{eq:initial_angle}). This guarantees that the neutrino self-interaction potential at the beginning of simulation is identical among all models. We assume that the reaction rate is proportional to $E_\nu^2$ emulating the energy dependence of neutral current scatterings (see Eqs.~\ref{eq:rho} and \ref{eq:rhob}).
To discuss multi-energy effects of collisions appropriately, we set the reaction rate so that the energy-integrated rate
becomes the same as that of E1$\_$HR model.
More specifically, we determine $R_0$ with a condition that the energy-averaged reaction rate for $\nu_e$,
\begin{align}
    \left<R_{ee}\right> = \frac{R(E_{\rm low})n_{\rm low}+R(E_{\rm high})n_{\rm high}}{n_\nu}, \label{eq:average}
\end{align}
becomes $R_{\rm HR}=1.25\times 10^{-4}\ {\rm cm^{-1}}$, which corresponds to the reaction rate used in E1$\_$HR model.
The resultant reaction rates at the energy of $E_{\rm low}$ and $E_{\rm high}$ are
$R(E_{\rm low}) = 2.50\times10^{-5}~{\rm cm^{-1}}$ and $R(E_{\rm high}) = 2.25\times10^{-4}~{\rm cm^{-1}}$, respectively, and the correspond $R_0$ is $10^{-4}\ {\rm cm^{-1}}$.
We name this model as a E2$\_$HR model.

The time evolution of $n_x$ in
E2$\_$HR model is depicted with a coral line in Figure~\ref{fig:rhoxx_scateffect}(b).
In the early phase, the time evolution of $n_x$ in E2$\_$HR model is almost identical to that of E1$\_$HR model.
After the time of $t \sim 10^{-7}$ s, differences between the two models become perceptible. The growth rate of $n_x$ in E2$\_$HR model becomes smaller than E1$\_$HR model, whereas the saturation amplitude of $n_x$ becomes higher. We have already witnessed the same trend when we compared the time evolution of $n_x$ between E1$\_$HR and E1$\_$LR models (see Section~\ref{sub:scat_effect}). Following the same argument, our result suggests that the frequency of collisions in E2$\_$HR model is reduced by multi-energy effects.

By comparing between the bottom-middle panel of Figure~\ref{fig:eln} and the left-top panel of Figure~\ref{fig:angle_tot_E2} (displaying $g_{ee}$ and $\bar{g}_{ee}$ as a function of neutrinos directional cosines in momentum space), the trend of weak collisions of E2$\_$HR model can also be seen.
The isotropization of $\nu_e$ is slower in E2$\_$HR model than E1$\_$HR model; consequently the $\delta$ in E2$\_$HR model also increases more slowly than that in the E1$\_$HR model, which is shown in Figure~\ref{fig:deg_asym}(b).
We also find that FFCs in E2$\_$HR model occur at the narrower angular range than in E1$\_$HR model, which is consistent with the trend found in the case of low reaction rate of energy-independent collisions.

One of the keys to understand the difference between the E1$\_$HR and E2$\_$HR models is rolls of collisions on high-energy neutrinos.
In the middle and right panels of Figure~\ref{fig:angle_tot_E2}, we display angular distributions of neutrinos at the energies of $E_{\rm low}$ and $E_{\rm high}$, respectively. 
We note that $g_{ee,{\rm low}}$ and $g_{ee,{\rm high}}$ are the energy-integrated angular distributions for low- and high-energy neutrinos; hence, they satisfy the condition of $g_{ee} = g_{ee,{\rm low}}+g_{ee,{\rm high}}$.
Since high-energy neutrinos experience collisions more frequently than those in E1$\_$HR model ($R(E_{\rm high})>R_{\rm HR}$), the angular distribution becomes more
isotropic.
As a result, the cancellation between in- and out-scatterings happens, which accounts for the effective reduction of collisions in E2$\_$HR model.
This feature can also be seen 
in the time evolution of $\delta$.
We display them for $E_{\rm low}$ and $E_{\rm high}$ as dashed and dash-dotted coral lines, respectively, in Figure~\ref{fig:deg_asym}(b).

To check if our interpretation, that the scattering-balances happening in high-energy neutrinos is responsible for the difference of FFCs between energy-dependent and independent collisions, is correct, we perform another simulation (E2$\_$HR$\_$f model),
in which the angular distribution of high-energy $\bar{\nu}_e$'s
is set to be isotropic at the beginning of the simulation. We also change the initial angular distribution of low-energy $\bar{\nu}_e$'s so as to be the same energy-integrated angular distributions of E2$\_$HR model, which guarantees that the neutrino self-interaction becomes identical to that of E2$\_$HR model.
If the cancellation between in- and out- scatterings is a key of multi-energy effects, FFC dynamics in E2$\_$HR$\_$f model further deviates from E1$\_$HR than the case of E2$\_$HR.

As shown in Figure~\ref{fig:rhoxx_scateffect}(b), the growth rate of $n_x$ for E2$\_$HR$\_$f model (purple) is slower but the saturation amplitude becomes higher than those in E2$\_$HR model (coral), that supports our interpretation. The time evolution of $\delta$ displayed in Figure~\ref{fig:deg_asym}(b) also provides another metric for this check.
As shown in the purple lines, $\delta$ is less energy-dependent up to the time of $\sim 2 \times 10^{-7}$~s for E2$\_$HR$\_$f model (purple), meanwhile the energy-dependence appears in the earlier phase for E2$\_$HR model (coral). This exhibits that effects of collisions in E2$\_$HR$\_$f model are more suppressed than E2$\_$HR model. In the late phase, on the other hand, the energy-integrated $\delta$ gradually increases with time, and eventually it becomes higher than that of E2$\_$HR model. We also display the angular distributions of $\nu_e$ and $\bar{\nu}_e$ for E2$\_$HR$\_$f model in bottom panels of Figure~\ref{fig:angle_tot_E2}, and we find that the time evolution is slower than E2$\_$HR model (displayed in the top panels of the same figure). All these results are consistent with our claim that the cancellation between in- and out-scatterings plays a primary role in the deviation of FFCs from those with energy-independent collisions.

\begin{figure}
\hspace*{-1.0cm}
\includegraphics[width=8.5cm]{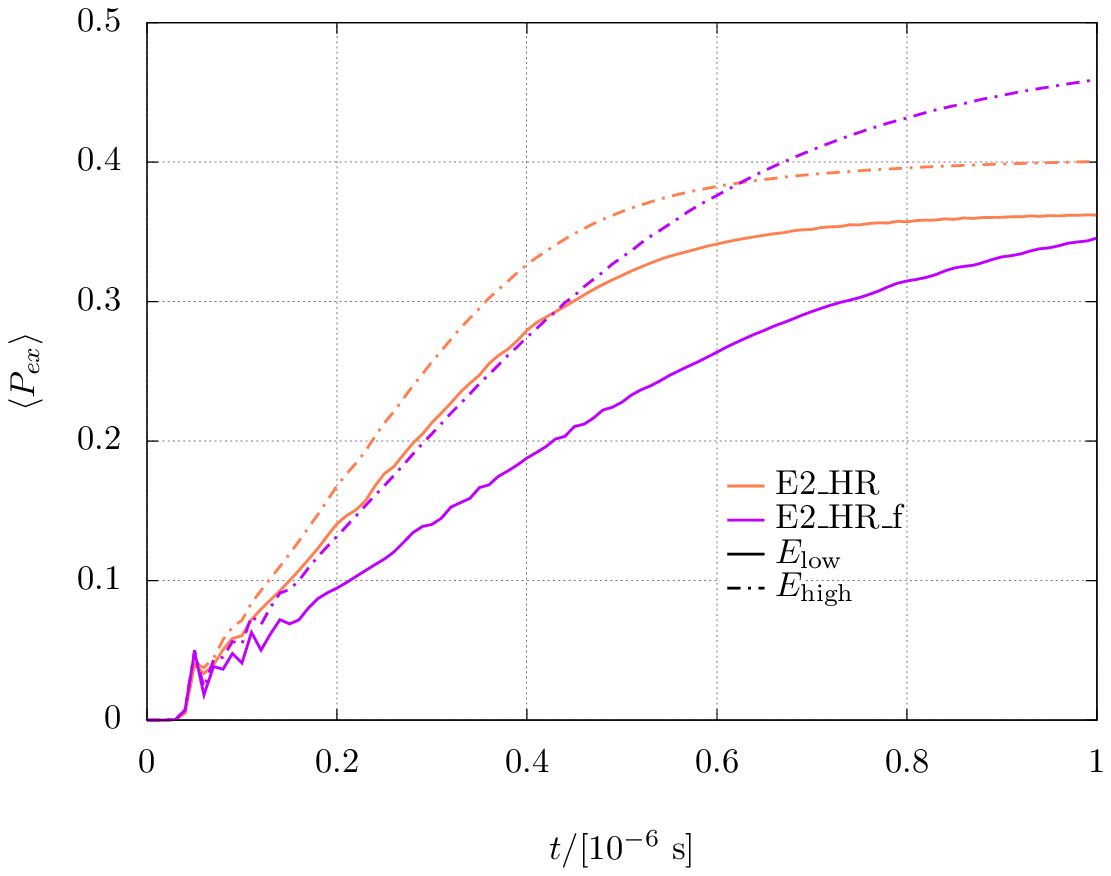}
\hspace*{-1.0cm}
\includegraphics[width=8.5cm]{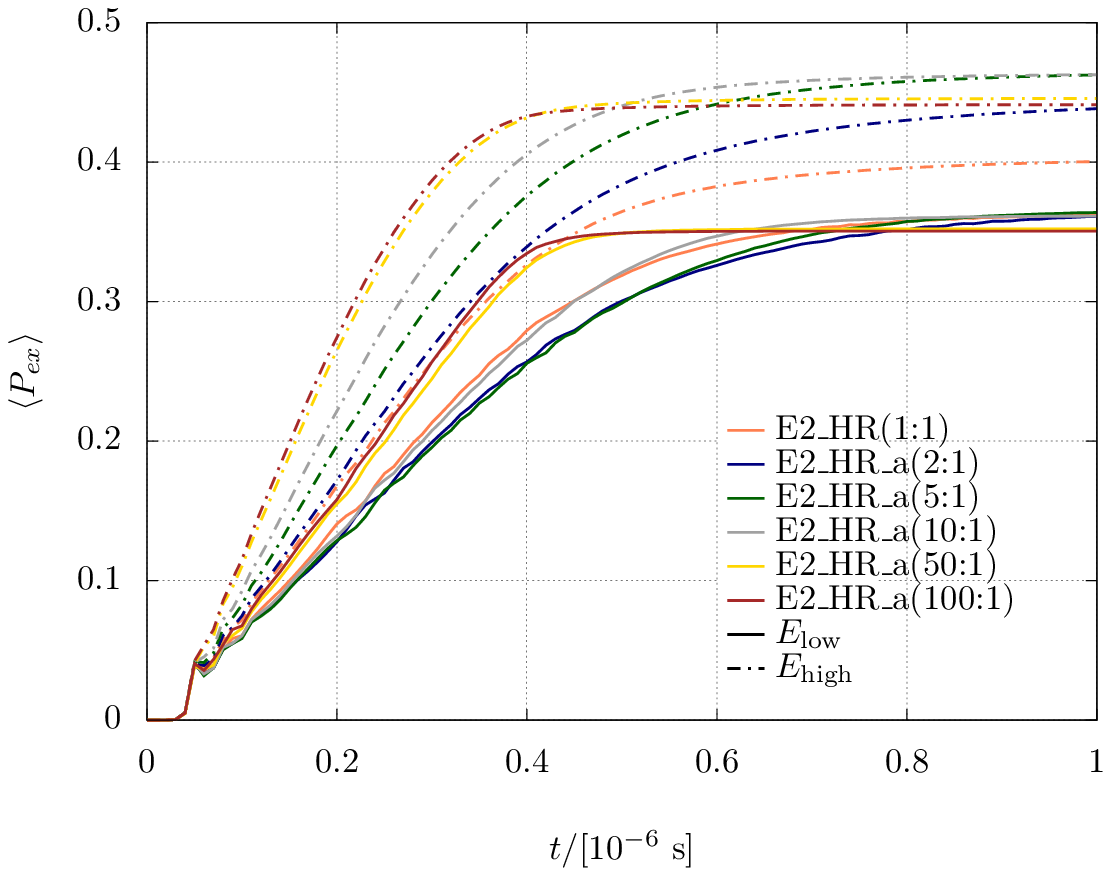}
\caption{\label{fig:pex} Time evolution of $\left<P_{ex}\right>$ for E2$\_$HR (top) and E2$\_$HR$\_$a models (bottom). Colors distinguish the models. Solid and dash-dotted lines denote the cases of low- and high-energy neutrinos, respectively.}
\end{figure}

To quantify the energy-dependent flavor conversions, we define a transition probability from $\nu_e$ to $\nu_x$ as,
\begin{eqnarray}
\left<P_{ex,i}\right> = 1-\frac{\int g_{ee,i} d\cos{\theta_\nu}}{\int g_{ee,i0} d\cos{\theta_\nu}},
\end{eqnarray}
with $i$= low or high.
We display them for E2$\_$HR and E2$\_$HR$\_$f models in the top panel of Figure~\ref{fig:pex}.
Solid and dash-dotted lines represent the cases of low- and high-energy neutrinos, respectively.
As shown in the figure, the transition probability is always higher in the case of high-energy neutrinos, which is due to effects of collisions; in fact the growth rate of FFCs becomes higher with increasing the reaction rate (see Section~\ref{sub:scat_effect}). It is worthy to note that flavor conversions in high energy region do not stop even if their angular distributions become nearly isotropic. 
In fact, the saturation time of $\left<P_{ex,i}\right>$ does not depend on energy. This exhibits that flavor conversions are driven by energy-integrated neutrino distributions, which is consistent with a property of FFCs.

\subsection{Energy-dependent collisions: non-flat neutrino-energy spectrum}  \label{sub:asym_effect}

\begin{table*} 
\begin{ruledtabular}
\begin{tabular}{lcccccccc}
&\multicolumn{1}{c}{$R_0/[10^{-4}{\rm cm^{-1}}]$} &
\multicolumn{1}{c}{$n_{\rm low}/n_\nu$} &
\multicolumn{1}{c}{$n_{\rm high}/n_\nu$} &
\multicolumn{1}{c}{$\beta$} &
\multicolumn{1}{c}{$R_{\rm low}/[10^{-4}{\rm cm^{-1}}]$} &
\multicolumn{1}{c}{$R_{\rm high}/[10^{-4}{\rm cm^{-1}}]$} &
\multicolumn{1}{c}{$\chi$}
\\
\colrule
E2$\_$HR(1:1) & 1.00 & 1/2 & 1/2 & 1 & 0.25 & 2.25 & 0.167\\
E2$\_$HR$\_$a(2:1) & 1.36 & 2/3 & 1/3 & 2 & 0.34 & 3.06 & 0.225 \\
E2$\_$HR$\_$a(5:1) & 2.14 & 5/6 & 1/6 & 5 & 0.54 & 4.82 & 0.263\\
E2$\_$HR$\_$a(10:1) & 2.89 & 10/11 & 1/11 & 10 & 0.72 & 6.50 & 0.245 \\
E2$\_$HR$\_$a(50:1) & 4.32 & 50/51 & 1/51 & 50 & 1.08 & 9.72 & 0.113 \\
E2$\_$HR$\_$a(100:1) & 4.63 & 100/101 & 1/101 & 100 & 1.16 & 10.4 & 0.078 \\
\end{tabular}
\end{ruledtabular}
\caption{Setups of E2$\_$HR and E2$\_$HR$\_$a models.}
\label{table:model}
\end{table*}

\begin{figure}
\hspace*{-1cm}
\includegraphics[width=9.2cm]{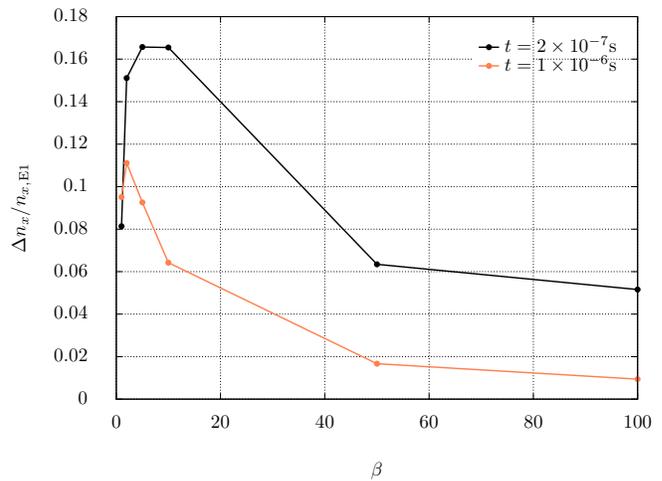}
\caption{\label{fig:rhoxx_asym} 
$\beta$ dependence of $\Delta n_x$ (normalized by $n_{x, {\rm E1}}$).
Results at two time snapshots are displayed:
$t=2\times10^{-7}$ (black) and $1\times10^{-6}$ s (coral). }
\end{figure}

It is an intriguing question how the above result (flat neutrino-energy spectrum) is altered in non-flat energy spectrum, i.e., $n_{\rm low}\ne n_{\rm high}$. 
Addressing the question is also important in studying for FFCs in SNe or BNSMs; in fact energy spectrum of neutrinos are non-monochromatic in these environments.
In this subsection, we study the case with non-flat neutrino energy spectrum, in which
we use E2$\_$HR model as a reference to determine the numerical setups. The models which we study in this section are referred to as E2$\_$HR$\_$a models hereafter.

We consider 6 models with $n_{\rm low}$:$n_{\rm high}$ = 1:1, 2:1, 5:1, 10:1, 50:1 and 100:1, satisfying
a condition of $n_{\rm low}+n_{\rm high} = n_\nu$, initially.
We note that the model with $n_{\rm low}=n_{\rm high}$ corresponds to E2$\_$HR model.
In the following discussions, we use the degree of neutrino-number-asymmetry ($\beta \equiv n_{\rm low}/n_{\rm high}$) to specify models.
As with the case of flat energy spectrum, neutrino energies are set to be $E_{\rm low}=10$ MeV and $E_{\rm high}=30$ MeV; we control the coefficient $R_0$ so that the average reaction rates become $R_{\rm HR}=1.25\times 10^{-4}\ {\rm cm^{-1}}$ (see Eq.~\ref{eq:average}).
We note that $R_0$ increases with $\beta$, since the dominance of low-energy neutrinos requires higher $R_0$ so as for $\left<R_{ee}\right>$ to be $R_{\rm HR}$.
The initial angular distributions of $g$'s are adopted from Eqs.~\ref{eq:initial_angle}, and the shapes are energy-independent.
The setup parameters are summarized in Table~\ref{table:model}.

We find that all models have essentially the same trend as that found in the flat spectrum; the time evolution of FFCs tends to deviate towards the case with lower reaction rate in energy-independent collisions.
More specifically, the $n_x$ saturation values and times in the asymmetric models are larger and later than those in E1$\_$HR model, respectively, which can be seen in Figure~\ref{fig:rhoxx_scateffect}(c) (see solid lines and cyan-dotted, where the former denotes E2$\_$HR$\_$a models and the latter is for E1$\_$HR model).
This strengthens our conclusion that the essence of multi-energy effects of collisions on FFCs is to reduce the reaction rate effectively due to the cancellation between in- and out-scatterings at high-energy neutrinos.

Although the overall trend is the same as the case of flat energy spectrum, the detailed feature of FFCs depends on the
spectrum.
To measure the sensitivity quantitatively, we use the absolute difference of $n_x$ from that of E1$\_$HR model, i.e., $\Delta n_x \equiv |n_{x} - n_{x, {\rm E1}}|$, where $n_{x, {\rm E1}}$ denotes the $n_x$ for E1$\_$HR model.
In Figure~\ref{fig:rhoxx_asym},
we display them (normalized by $n_{x, {\rm E1}}$) as a function of $\beta$ at $t=2\times10^{-7}$ and $1\times10^{-6}$ s.
As shown in the figure, $\Delta n_x$ is a non-monotonic dependence on $\beta$. 
This can be understood by considering the two opposite limits: $n_{\rm low} \rightarrow 0$ ($\beta \rightarrow 0$) and $n_{\rm high} \rightarrow 0$ ($\beta \rightarrow \infty$). Both cases are identical to E1$\_$HR model, implying that $\Delta n_x$ should vary non-monotonically with $\beta$.
We also find that $\Delta n_x$ has a single peak, which is located at $1 \lesssim \beta \lesssim 10$.
Below, we attempt to understand the reason why $\Delta n_x$ becomes the largest at $\beta = \mathcal{O}(1)$.

There are mainly two important elements to understand this trend. First, multi-energy effects of collisions should hinge on the dispersion of neutrino energy spectrum. One can expect that they tend to be larger for neutrinos with a broad energy spectrum. On the other hand, the dispersion of neutrino energy spectrum is not enough to quantify multi-energy effects of collisions. In fact, there are no effects if reaction rates do not have the energy dependence (see in Section~\ref{sub:scat_effect}). We develop a new diagnostics to quantify multi-energy effects, which satisfies above considerations. As shown below, this diagnostics suggests that multi-energy effects become the highest at $\beta = \mathcal{O}(1)$.

We quantify the impact of multi-energy effects on FFCs with a new variable, $\chi$, in which we compute the difference between the average energy of neutrinos and reaction-weighted one. More specifically, $\chi$ is defined as,
\begin{eqnarray}
\chi = \left|\frac{\left<E_\nu\right>-\left<RE_\nu\right>}{\left<E_\nu\right>+\left<RE_\nu\right>}\right|,
\label{eq:chi}
\end{eqnarray}
where
\begin{eqnarray}
\left<E_\nu\right> &=& \frac{\int d^3\vec{p} E_\nu\rho }{\int d^3\vec{p} \rho }, \label{eq:avenuEne} \\
\left<RE_\nu\right> &=& \frac{\int d^3\vec{p} R E_\nu\rho }{\int d^3\vec{p} R\rho }.
\label{eq:avenuEne_Rweight}
\end{eqnarray}
For monochromatic neutrinos, we obtain $\left<E_\nu\right> = \left<RE_\nu\right>$, leading to $\chi=0$, which is consistent with no multi-energy effects. In the case with non-monochromatic energy spectrum of neutrinos but no energy dependence in reaction rates, we again obtain $\chi=0$, guaranteeing no multi-energy effects. For energy-dependent reactions, $\chi$ has, in general, a non-zero value, and it becomes higher for broader energy spectrum. This is in line with what we expect in multi-energy effects.

We apply the $\chi$ diagnostics to our E2 models. The $\left<E_\nu\right>$ and $\left<RE_\nu\right>$ can be written as
\begin{eqnarray}
\left<E_\nu\right> &=& \frac{E_{\rm low}\beta+E_{\rm high}}{\beta + 1}, \\
\left<RE_\nu\right> &=& \frac{E_{\rm low}^3\beta + E_{\rm high}^3}{E_{\rm low}^2\beta + E_{\rm high}^2}, \label{eq:REnu_beta}
\end{eqnarray}
and we can obtain $\chi$ from Eq.~\ref{eq:chi}. $\chi$ for all E2 models are listed in Table~\ref{table:model}. We find that $\chi$ becomes maximum around $\beta \sim 5$, which is consistent with the peak at $\beta\sim\mathcal{O}(1)$ in Figure~\ref{fig:rhoxx_asym}, exhibiting that $\chi$ captures the trend of multi-energy effects of collision qualitatively.

It should be noted that the $\chi$ diagnostics is not capable of determining the exact value of $\beta$, which has the maximum of $\Delta n_x$. In fact, $\Delta n_x$ at $t=1\times10^{-6}$~s becomes the largest at $\beta=2$ among our models, which is different from the $\chi$ diagnostics. This is due to the fact that there are a number of complexities that affect $\Delta n_x$ through non-linear interactions between FFCs and collisions. This can be seen in the time evolution of $\delta$, which is displayed in Figure~\ref{fig:deg_asym}(c).
In the figure, we display results of $\beta=2$ and 5 models in navy and dark-green lines, respectively. 
As shown by the solid lines, $\delta$ for the energy-integrated angular distributions in the $\beta=5$ model is slightly lower than that in the $\beta=2$ one until $t \sim 5 \times 10^{-7}$~s. This indicates that the deviation from E1 model (the difference from the cyan line in Figure~\ref{fig:deg_asym}(c)) is higher for $\beta=5$ model than $\beta=2$ one, which is consistent with the $\chi$ diagnostics. In the late phase ($t \gtrsim 5 \times 10^{-7}$~s), however, the time evolution of $\delta$ becomes very complex; $\delta$ in the $\beta=5$ model becomes higher than the $\beta=2$ one in $ 5 \times 10^{-7} \lesssim t \lesssim 8 \times 10^{-7}$~s, but the order is again reversed after $t\sim 8 \times 10^{-7}$~s. To capture these temporal features, more elaborate diagnostics needs to be developed\footnote{The slow but long-time increase of $\delta$ in the $\beta=2$ model displayed in Figure~\ref{fig:deg_asym}(c) may be interpreted through $R_{\rm low}$.
Since $R_{\rm low}$ of $\beta=2$ model is lower than that of $\beta=5$ one, the life time of FFCs in the $\beta=2$ model becomes longer and the total amount of flavor conversions becomes higher than those in $\beta=5$ model (see also Section~\ref{sub:scat_effect}).}.
It should be emphasized, however, that the deviation of the peak $\beta$ is a factor of a few in the 
 $\chi$ diagnostics, suggesting that the diagnostics is still informative to exhibit the dependence of $\Delta n_x$ on $\beta$ qualitatively.

Finally, we show the time evolution of $\left<P_{ex}\right>$ in the bottom panel of Figure~\ref{fig:pex}. 
The figure illustrates the impact of energy dependence of collisions on FFCs in models with non-flat energy spectrum. This leads us to a robust conclusion that multi-energy effects of collisions induce energy-dependent features in FFCs, and the mixing degree sensitively depends on neutrino energy spectrum.

\subsection{$\chi$ diagnostics in three energy meshes}  \label{sub:threemeshes}

One of the virtues of $\chi$ diagnostics is that it can be applied to arbitrary energy spectra of neutrinos (see Eqs.~\ref{eq:chi}-\ref{eq:avenuEne_Rweight}). Although it is useful for the study of SNe and BNSMs, we need to assess how well the diagnostics can quantify multi-energy effects of collisions on FFCs in cases with $N_{\varepsilon} \geq 3$, where $N_{\varepsilon}$ denotes the number of energy meshes. In this section, we run another FFC simulation with three energy meshes to check this concern.

The initial condition is set in the same manner as in the case of E2 models.
Neutrino energies are assumed to be 6.67, 20.0 and 33.3 MeV, and the energy spectrum is flat.
The initial angular distributions of $g$'s are adopted from Eqs.~\ref{eq:initial_angle}, and the shapes are energy-independent. We take $R_0=9.64\times10^{-5}\ {\rm cm^{-1}}$ so that the average reaction rates become $R_{\rm HR}=1.25\times10^{-4}\ {\rm cm^{-1}}$ (E3$\_$HR). In E3$\_$HR model, $\chi$ becomes 0.186, which is similar to that of E2$\_$HR model ($\chi$=0.167).

\begin{figure}
\hspace*{-1cm}
\includegraphics[width=9.2cm]{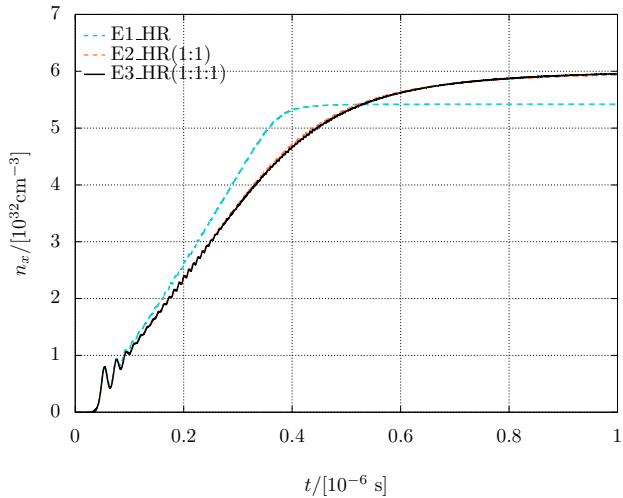}
\caption{\label{fig:E3} 
Time evolution of $n_x$ for E3$\_$HR model. }
\end{figure}

In Figure~\ref{fig:E3}, we compare the time evolution of $n_x$ for E3$\_$HR model to those in E1$\_$HR and E2$\_$HR models. We find that E3$\_$HR model is almost identical to E2$\_$HR one. It is consistent with the $\chi$ diagnostics, lending confidence to its applicability in the case with $N_{\varepsilon} = 3$.

It should be stressed that more extensive tests are obviously needed to assess the applicability of the $\chi$ diagnostics to more general cases such as neutrino radiation fields in SNe and BNSMs. We leave this important task for future works.

\section{Summary and discussions} \label{sec:summary}
Recent studies have indicated that fast neutrino flavor conversion (FFC), one of neutrino oscillations induced by neutrino self-interaction, occurs in SNe or BNSMs.
However, there is still little known about the non-linear dynamics. One of unresolved issues is the interplay between FFCs and neutrino-matter collisions, which is the main focus of this paper.

FFCs are essentially energy-independent; hence, almost all previous studies have taken the monochromatic or energy-integrated assumption.
On the other hand, neutrino-matter collisions are in general energy-dependent, suggesting that the energy-integrated assumption may discard some potentially important effects of collisions on FFCs.

Motivated by this fact, we perform
dynamical simulations of FFCs with iso-energetic scatterings by our QKE-MC solver \cite{kato2021}.
We start with revisiting the case of energy-independent collisions in Section~\ref{sub:scat_effect}.
The $n_x$ evolution without collisions has a periodic feature with a constant amplitude, while $n_x$ increases with small oscillations, and eventually it reaches the saturation at a certain time in the case with collisions.
There are also clear differences in angular distributions of neutrinos between with and without collisions.
In the case without collisions, FFCs occur in a very narrow angular region, which is due to the fact that the ELN-XLN angular crossing point is almost constant with time. In cases with collisions, on the other hand, the crossing point becomes dynamical due to effects of collisions; consequently, FFCs occur in wider angular regions than the case without collisions.
We also find that FFCs in higher-reaction rate model grow faster, but they are saturated at earlier time. As a result, the saturation value of $n_x$ tends to be smaller with increasing the reaction rate.

In Section~\ref{sub:multi_effect}, we investigate effects of energy-dependent collisions on FFCs
under the assumption of flat neutrino-energy spectrum.
To capture the qualitative trend, we study the impact with two energy meshes.
The energy dependence of reaction rate is proportional to $E_\nu^2$, emulating neutrino-nucleon scatterings $(R\propto E_\nu^2)$.
Our simulations show the clear difference between energy-independent and dependent collisions; the latter tends to be similar evolution as that with a lower reaction rate in the energy-independent collision.
This is because high-energy neutrinos experience collisions more frequently, implying that their angular distributions become more isotropic. As a result, in- and out-scatterings are cancelled out each other, which accounts for reducing the number of collisions effectively.
We also find that the transition probability $\left<P_{ex}\right>$ depends on neutrino energy, exhibiting the importance of multi-energy treatment in studying FFCs with collisions.

In Section~\ref{sub:asym_effect}, we discuss the case of non-flat neutrino-energy spectrum.
The overall trend is in line with the case of flat energy-spectrum; in fact, FFCs slowly evolve with time, but its lifetime becomes longer due to the effective reduction in the number of collisions. On the other hand, the detailed feature of FFCs depends on the energy-spectrum, and we find that the large deviation from the case with energy-independent collisions happens for $\beta (\equiv n_{\rm low}/n_{\rm high}) \sim \mathcal{O}(1)$. To understand the trend of deviation, we propose a new diagnostics with $\chi$ (see Eqs.~\ref{eq:chi}-\ref{eq:avenuEne_Rweight}), in which we quantify multi-energy effects of collisions on FFCs by the difference between the average-energy of neutrinos and reaction-weighted one. Although the $\chi$ diagnostics is not capable of capturing all features of multi-energy effects (such as temporal features), the diagnostics is in reasonable agreement with the numerical results. In Section~\ref{sub:threemeshes}, we further assess the capability of the diagnostics by comparing to another FFC simulation with three energy meshes, and we confirm that the numerical results are consistent with the diagnostics.

Another thing we do notice from the present study is that the transition probability of flavor conversion is energy-dependent.
This also sensitively depends on the neutrino-energy spectrum. 
Our finding opens a new path to create energy-dependent flavor conversions driven by FFCs, which should be distinct from the slow mode.

Obviously, we imposed many assumptions and simplifications in the present study. Although these conditions are useful to highlight effects of energy-dependent collisions on FFCs, they should be relaxed for more realistic situations.
The top priority would be to get rid of homogeneous condition. As is well known, the homogeneous condition restricts the dynamics of FFCs \cite{abbar2019,martin2020,sherwood2021,bhattacharyya2020}, and more importantly, it also faces a self-consistency issue to consider effects of collisions on flavor conversions \cite{lucas2022}. It is also an intriguing question how other collision terms,  (emissions, absorptions, inelastic scatterings, and pair processes) create energy-dependent features in FFCs.
We leave the investigation on these remaining issues to future works.

It should be stressed, however, that our main conclusion, emergence of energy-dependent FFCs induced by collisions, would not be changed in more realistic situations. Hence, it is particularly interesting to delve into FFCs in high-energy neutrinos with energy-dependent collisions; the reason is as follows. In classical neutrino transport, those neutrinos tend to be in thermal/chemical equilibrium with matter due to strong neutrino-matter coupling. Since the equilibrium state hinges on a neutrino flavor, the competition between FFCs and neutrino-matter interactions would happen. This competition may induce unexpected phenomena in both neutrino-flavor conversion and fluid dynamics. Addressing this issue 
would be important to gauge the sensitivity of FFCs to SNe and BNSMs. We also leave this exploration to future works.

\begin{acknowledgements}
We are grateful to Lucas Johns and Masamichi Zaizen for useful comments and discussions. C. K. is supported by JSPS KAKENHI Grant Numbers JP20K14457 and JP22H04577.
The numerical calculations were carried out on Yukawa-21 at YITP in Kyoto University and Cray XC50 at Center for Computational Astrophysics, National Astronomical Observatory of Japan.
\end{acknowledgements}

\bibliography{MCenergy}

\begin{thebibliography}{55}%
\makeatletter
\providecommand \@ifxundefined [1]{%
 \@ifx{#1\undefined}
}%
\providecommand \@ifnum [1]{%
 \ifnum #1\expandafter \@firstoftwo
 \else \expandafter \@secondoftwo
 \fi
}%
\providecommand \@ifx [1]{%
 \ifx #1\expandafter \@firstoftwo
 \else \expandafter \@secondoftwo
 \fi
}%
\providecommand \natexlab [1]{#1}%
\providecommand \enquote  [1]{``#1''}%
\providecommand \bibnamefont  [1]{#1}%
\providecommand \bibfnamefont [1]{#1}%
\providecommand \citenamefont [1]{#1}%
\providecommand \href@noop [0]{\@secondoftwo}%
\providecommand \href [0]{\begingroup \@sanitize@url \@href}%
\providecommand \@href[1]{\@@startlink{#1}\@@href}%
\providecommand \@@href[1]{\endgroup#1\@@endlink}%
\providecommand \@sanitize@url [0]{\catcode `\\12\catcode `\$12\catcode
  `\&12\catcode `\#12\catcode `\^12\catcode `\_12\catcode `\%12\relax}%
\providecommand \@@startlink[1]{}%
\providecommand \@@endlink[0]{}%
\providecommand \url  [0]{\begingroup\@sanitize@url \@url }%
\providecommand \@url [1]{\endgroup\@href {#1}{\urlprefix }}%
\providecommand \urlprefix  [0]{URL }%
\providecommand \Eprint [0]{\href }%
\providecommand \doibase [0]{https://doi.org/}%
\providecommand \selectlanguage [0]{\@gobble}%
\providecommand \bibinfo  [0]{\@secondoftwo}%
\providecommand \bibfield  [0]{\@secondoftwo}%
\providecommand \translation [1]{[#1]}%
\providecommand \BibitemOpen [0]{}%
\providecommand \bibitemStop [0]{}%
\providecommand \bibitemNoStop [0]{.\EOS\space}%
\providecommand \EOS [0]{\spacefactor3000\relax}%
\providecommand \BibitemShut  [1]{\csname bibitem#1\endcsname}%
\let\auto@bib@innerbib\@empty
\bibitem [{\citenamefont {Samuel}(1993)}]{samuel1993}%
  \BibitemOpen
  \bibfield  {author} {\bibinfo {author} {\bibfnamefont {S.}~\bibnamefont
  {Samuel}},\ }\bibfield  {title} {\bibinfo {title} {Neutrino oscillations in
  dense neutrino gases},\ }\href {https://doi.org/10.1103/PhysRevD.48.1462}
  {\bibfield  {journal} {\bibinfo  {journal} {Phys. Rev. D}\ }\textbf {\bibinfo
  {volume} {48}},\ \bibinfo {pages} {1462} (\bibinfo {year}
  {1993})}\BibitemShut {NoStop}%
\bibitem [{\citenamefont {{Sigl}}\ and\ \citenamefont
  {{Raffelt}}(1993)}]{sigl1993}%
  \BibitemOpen
  \bibfield  {author} {\bibinfo {author} {\bibfnamefont {G.}~\bibnamefont
  {{Sigl}}}\ and\ \bibinfo {author} {\bibfnamefont {G.}~\bibnamefont
  {{Raffelt}}},\ }\bibfield  {title} {\bibinfo {title} {{General kinetic
  description of relativistic mixed neutrinos}},\ }\href
  {https://doi.org/10.1016/0550-3213(93)90175-O} {\bibfield  {journal}
  {\bibinfo  {journal} {Nuclear Physics B}\ }\textbf {\bibinfo {volume}
  {406}},\ \bibinfo {pages} {423} (\bibinfo {year} {1993})}\BibitemShut
  {NoStop}%
\bibitem [{\citenamefont {Sigl}(1995)}]{sigl1995}%
  \BibitemOpen
  \bibfield  {author} {\bibinfo {author} {\bibfnamefont {G.}~\bibnamefont
  {Sigl}},\ }\bibfield  {title} {\bibinfo {title} {Neutrino mixing constraints
  and supernova nucleosynthesis},\ }\href
  {https://doi.org/10.1103/PhysRevD.51.4035} {\bibfield  {journal} {\bibinfo
  {journal} {Phys. Rev. D}\ }\textbf {\bibinfo {volume} {51}},\ \bibinfo
  {pages} {4035} (\bibinfo {year} {1995})}\BibitemShut {NoStop}%
\bibitem [{\citenamefont {Sawyer}(2005)}]{sawyer2005}%
  \BibitemOpen
  \bibfield  {author} {\bibinfo {author} {\bibfnamefont {R.~F.}\ \bibnamefont
  {Sawyer}},\ }\bibfield  {title} {\bibinfo {title} {Speed-up of neutrino
  transformations in a supernova environment},\ }\href
  {https://doi.org/10.1103/PhysRevD.72.045003} {\bibfield  {journal} {\bibinfo
  {journal} {Phys. Rev. D}\ }\textbf {\bibinfo {volume} {72}},\ \bibinfo
  {pages} {045003} (\bibinfo {year} {2005})}\BibitemShut {NoStop}%
\bibitem [{\citenamefont {{Morinaga}}(2022)}]{morinaga2021}%
  \BibitemOpen
  \bibfield  {author} {\bibinfo {author} {\bibfnamefont {T.}~\bibnamefont
  {{Morinaga}}},\ }\bibfield  {title} {\bibinfo {title} {{Fast neutrino flavor
  instability and neutrino flavor lepton number crossings}},\ }\href
  {https://doi.org/10.1103/PhysRevD.105.L101301} {\bibfield  {journal}
  {\bibinfo  {journal} {\prd}\ }\textbf {\bibinfo {volume} {105}},\ \bibinfo
  {eid} {L101301} (\bibinfo {year} {2022})},\ \Eprint
  {https://arxiv.org/abs/2103.15267} {arXiv:2103.15267 [hep-ph]} \BibitemShut
  {NoStop}%
\bibitem [{\citenamefont {{Dasgupta}}\ \emph {et~al.}(2017)\citenamefont
  {{Dasgupta}}, \citenamefont {{Mirizzi}},\ and\ \citenamefont
  {{Sen}}}]{dasgupta2017}%
  \BibitemOpen
  \bibfield  {author} {\bibinfo {author} {\bibfnamefont {B.}~\bibnamefont
  {{Dasgupta}}}, \bibinfo {author} {\bibfnamefont {A.}~\bibnamefont
  {{Mirizzi}}},\ and\ \bibinfo {author} {\bibfnamefont {M.}~\bibnamefont
  {{Sen}}},\ }\bibfield  {title} {\bibinfo {title} {{Fast neutrino flavor
  conversions near the supernova core with realistic flavor-dependent angular
  distributions}},\ }\href {https://doi.org/10.1088/1475-7516/2017/02/019}
  {\bibfield  {journal} {\bibinfo  {journal} {\jcap}\ }\textbf {\bibinfo
  {volume} {2017}},\ \bibinfo {eid} {019} (\bibinfo {year} {2017})},\ \Eprint
  {https://arxiv.org/abs/1609.00528} {arXiv:1609.00528 [hep-ph]} \BibitemShut
  {NoStop}%
\bibitem [{\citenamefont {{Tamborra}}\ \emph {et~al.}(2017)\citenamefont
  {{Tamborra}}, \citenamefont {{H{\"u}depohl}}, \citenamefont {{Raffelt}},\
  and\ \citenamefont {{Janka}}}]{tamborra2017}%
  \BibitemOpen
  \bibfield  {author} {\bibinfo {author} {\bibfnamefont {I.}~\bibnamefont
  {{Tamborra}}}, \bibinfo {author} {\bibfnamefont {L.}~\bibnamefont
  {{H{\"u}depohl}}}, \bibinfo {author} {\bibfnamefont {G.~G.}\ \bibnamefont
  {{Raffelt}}},\ and\ \bibinfo {author} {\bibfnamefont {H.-T.}\ \bibnamefont
  {{Janka}}},\ }\bibfield  {title} {\bibinfo {title} {{Flavor-dependent
  Neutrino Angular Distribution in Core-collapse Supernovae}},\ }\href
  {https://doi.org/10.3847/1538-4357/aa6a18} {\bibfield  {journal} {\bibinfo
  {journal} {\apj}\ }\textbf {\bibinfo {volume} {839}},\ \bibinfo {eid} {132}
  (\bibinfo {year} {2017})},\ \Eprint {https://arxiv.org/abs/1702.00060}
  {arXiv:1702.00060 [astro-ph.HE]} \BibitemShut {NoStop}%
\bibitem [{\citenamefont {Wu}\ \emph {et~al.}(2017)\citenamefont {Wu},
  \citenamefont {Tamborra}, \citenamefont {Just},\ and\ \citenamefont
  {Janka}}]{wu2017}%
  \BibitemOpen
  \bibfield  {author} {\bibinfo {author} {\bibfnamefont {M.-R.}\ \bibnamefont
  {Wu}}, \bibinfo {author} {\bibfnamefont {I.}~\bibnamefont {Tamborra}},
  \bibinfo {author} {\bibfnamefont {O.}~\bibnamefont {Just}},\ and\ \bibinfo
  {author} {\bibfnamefont {H.-T.}\ \bibnamefont {Janka}},\ }\bibfield  {title}
  {\bibinfo {title} {Imprints of neutrino-pair flavor conversions on
  nucleosynthesis in ejecta from neutron-star merger remnants},\ }\href
  {https://doi.org/10.1103/PhysRevD.96.123015} {\bibfield  {journal} {\bibinfo
  {journal} {Phys. Rev. D}\ }\textbf {\bibinfo {volume} {96}},\ \bibinfo
  {pages} {123015} (\bibinfo {year} {2017})}\BibitemShut {NoStop}%
\bibitem [{\citenamefont {Dasgupta}\ \emph {et~al.}(2018)\citenamefont
  {Dasgupta}, \citenamefont {Mirizzi},\ and\ \citenamefont
  {Sen}}]{dasgupta2018}%
  \BibitemOpen
  \bibfield  {author} {\bibinfo {author} {\bibfnamefont {B.}~\bibnamefont
  {Dasgupta}}, \bibinfo {author} {\bibfnamefont {A.}~\bibnamefont {Mirizzi}},\
  and\ \bibinfo {author} {\bibfnamefont {M.}~\bibnamefont {Sen}},\ }\bibfield
  {title} {\bibinfo {title} {Simple method of diagnosing fast flavor
  conversions of supernova neutrinos},\ }\href
  {https://doi.org/10.1103/PhysRevD.98.103001} {\bibfield  {journal} {\bibinfo
  {journal} {Phys. Rev. D}\ }\textbf {\bibinfo {volume} {98}},\ \bibinfo
  {pages} {103001} (\bibinfo {year} {2018})}\BibitemShut {NoStop}%
\bibitem [{\citenamefont {{Abbar}}\ \emph {et~al.}(2019)\citenamefont
  {{Abbar}}, \citenamefont {{Duan}}, \citenamefont {{Sumiyoshi}}, \citenamefont
  {{Takiwaki}},\ and\ \citenamefont {{Volpe}}}]{abbar2019b}%
  \BibitemOpen
  \bibfield  {author} {\bibinfo {author} {\bibfnamefont {S.}~\bibnamefont
  {{Abbar}}}, \bibinfo {author} {\bibfnamefont {H.}~\bibnamefont {{Duan}}},
  \bibinfo {author} {\bibfnamefont {K.}~\bibnamefont {{Sumiyoshi}}}, \bibinfo
  {author} {\bibfnamefont {T.}~\bibnamefont {{Takiwaki}}},\ and\ \bibinfo
  {author} {\bibfnamefont {M.~C.}\ \bibnamefont {{Volpe}}},\ }\bibfield
  {title} {\bibinfo {title} {{On the occurrence of fast neutrino flavor
  conversions in multidimensional supernova models}},\ }\href
  {https://doi.org/10.1103/PhysRevD.100.043004} {\bibfield  {journal} {\bibinfo
   {journal} {\prd}\ }\textbf {\bibinfo {volume} {100}},\ \bibinfo {eid}
  {043004} (\bibinfo {year} {2019})},\ \Eprint
  {https://arxiv.org/abs/1812.06883} {arXiv:1812.06883 [astro-ph.HE]}
  \BibitemShut {NoStop}%
\bibitem [{\citenamefont {{Nagakura}}\ \emph {et~al.}(2019)\citenamefont
  {{Nagakura}}, \citenamefont {{Morinaga}}, \citenamefont {{Kato}},\ and\
  \citenamefont {{Yamada}}}]{nagakura2019}%
  \BibitemOpen
  \bibfield  {author} {\bibinfo {author} {\bibfnamefont {H.}~\bibnamefont
  {{Nagakura}}}, \bibinfo {author} {\bibfnamefont {T.}~\bibnamefont
  {{Morinaga}}}, \bibinfo {author} {\bibfnamefont {C.}~\bibnamefont {{Kato}}},\
  and\ \bibinfo {author} {\bibfnamefont {S.}~\bibnamefont {{Yamada}}},\
  }\bibfield  {title} {\bibinfo {title} {{Fast-pairwise Collective Neutrino
  Oscillations Associated with Asymmetric Neutrino Emissions in Core-collapse
  Supernovae}},\ }\href {https://doi.org/10.3847/1538-4357/ab4cf2} {\bibfield
  {journal} {\bibinfo  {journal} {\apj}\ }\textbf {\bibinfo {volume} {886}},\
  \bibinfo {eid} {139} (\bibinfo {year} {2019})},\ \Eprint
  {https://arxiv.org/abs/1910.04288} {arXiv:1910.04288 [astro-ph.HE]}
  \BibitemShut {NoStop}%
\bibitem [{\citenamefont {{Delfan Azari}}\ \emph {et~al.}(2019)\citenamefont
  {{Delfan Azari}}, \citenamefont {{Yamada}}, \citenamefont {{Morinaga}},
  \citenamefont {{Iwakami}}, \citenamefont {{Okawa}}, \citenamefont
  {{Nagakura}},\ and\ \citenamefont {{Sumiyoshi}}}]{milad2019}%
  \BibitemOpen
  \bibfield  {author} {\bibinfo {author} {\bibfnamefont {M.}~\bibnamefont
  {{Delfan Azari}}}, \bibinfo {author} {\bibfnamefont {S.}~\bibnamefont
  {{Yamada}}}, \bibinfo {author} {\bibfnamefont {T.}~\bibnamefont
  {{Morinaga}}}, \bibinfo {author} {\bibfnamefont {W.}~\bibnamefont
  {{Iwakami}}}, \bibinfo {author} {\bibfnamefont {H.}~\bibnamefont {{Okawa}}},
  \bibinfo {author} {\bibfnamefont {H.}~\bibnamefont {{Nagakura}}},\ and\
  \bibinfo {author} {\bibfnamefont {K.}~\bibnamefont {{Sumiyoshi}}},\
  }\bibfield  {title} {\bibinfo {title} {{Linear analysis of fast-pairwise
  collective neutrino oscillations in core-collapse supernovae based on the
  results of Boltzmann simulations}},\ }\href
  {https://doi.org/10.1103/PhysRevD.99.103011} {\bibfield  {journal} {\bibinfo
  {journal} {\prd}\ }\textbf {\bibinfo {volume} {99}},\ \bibinfo {eid} {103011}
  (\bibinfo {year} {2019})},\ \Eprint {https://arxiv.org/abs/1902.07467}
  {arXiv:1902.07467 [astro-ph.HE]} \BibitemShut {NoStop}%
\bibitem [{\citenamefont {Glas}\ \emph {et~al.}(2020)\citenamefont {Glas},
  \citenamefont {Janka}, \citenamefont {Capozzi}, \citenamefont {Sen},
  \citenamefont {Dasgupta}, \citenamefont {Mirizzi},\ and\ \citenamefont
  {Sigl}}]{glas2020}%
  \BibitemOpen
  \bibfield  {author} {\bibinfo {author} {\bibfnamefont {R.}~\bibnamefont
  {Glas}}, \bibinfo {author} {\bibfnamefont {H.-T.}\ \bibnamefont {Janka}},
  \bibinfo {author} {\bibfnamefont {F.}~\bibnamefont {Capozzi}}, \bibinfo
  {author} {\bibfnamefont {M.}~\bibnamefont {Sen}}, \bibinfo {author}
  {\bibfnamefont {B.}~\bibnamefont {Dasgupta}}, \bibinfo {author}
  {\bibfnamefont {A.}~\bibnamefont {Mirizzi}},\ and\ \bibinfo {author}
  {\bibfnamefont {G.}~\bibnamefont {Sigl}},\ }\bibfield  {title} {\bibinfo
  {title} {Fast neutrino flavor instability in the neutron-star convection
  layer of three-dimensional supernova models},\ }\href
  {https://doi.org/10.1103/PhysRevD.101.063001} {\bibfield  {journal} {\bibinfo
   {journal} {Phys. Rev. D}\ }\textbf {\bibinfo {volume} {101}},\ \bibinfo
  {pages} {063001} (\bibinfo {year} {2020})}\BibitemShut {NoStop}%
\bibitem [{\citenamefont {Abbar}\ \emph {et~al.}(2020)\citenamefont {Abbar},
  \citenamefont {Duan}, \citenamefont {Sumiyoshi}, \citenamefont {Takiwaki},\
  and\ \citenamefont {Volpe}}]{abbar2020}%
  \BibitemOpen
  \bibfield  {author} {\bibinfo {author} {\bibfnamefont {S.}~\bibnamefont
  {Abbar}}, \bibinfo {author} {\bibfnamefont {H.}~\bibnamefont {Duan}},
  \bibinfo {author} {\bibfnamefont {K.}~\bibnamefont {Sumiyoshi}}, \bibinfo
  {author} {\bibfnamefont {T.}~\bibnamefont {Takiwaki}},\ and\ \bibinfo
  {author} {\bibfnamefont {M.~C.}\ \bibnamefont {Volpe}},\ }\bibfield  {title}
  {\bibinfo {title} {Fast neutrino flavor conversion modes in multidimensional
  core-collapse supernova models: The role of the asymmetric neutrino
  distributions},\ }\href {https://doi.org/10.1103/PhysRevD.101.043016}
  {\bibfield  {journal} {\bibinfo  {journal} {Phys. Rev. D}\ }\textbf {\bibinfo
  {volume} {101}},\ \bibinfo {pages} {043016} (\bibinfo {year}
  {2020})}\BibitemShut {NoStop}%
\bibitem [{\citenamefont {{Abbar}}(2020)}]{abbar2020b}%
  \BibitemOpen
  \bibfield  {author} {\bibinfo {author} {\bibfnamefont {S.}~\bibnamefont
  {{Abbar}}},\ }\bibfield  {title} {\bibinfo {title} {{Searching for fast
  neutrino flavor conversion modes in core-collapse supernova simulations}},\
  }\href {https://doi.org/10.1088/1475-7516/2020/05/027} {\bibfield  {journal}
  {\bibinfo  {journal} {\jcap}\ }\textbf {\bibinfo {volume} {2020}},\ \bibinfo
  {eid} {027} (\bibinfo {year} {2020})},\ \Eprint
  {https://arxiv.org/abs/2003.00969} {arXiv:2003.00969 [astro-ph.HE]}
  \BibitemShut {NoStop}%
\bibitem [{\citenamefont {{Morinaga}}\ \emph {et~al.}(2020)\citenamefont
  {{Morinaga}}, \citenamefont {{Nagakura}}, \citenamefont {{Kato}},\ and\
  \citenamefont {{Yamada}}}]{morinaga2020}%
  \BibitemOpen
  \bibfield  {author} {\bibinfo {author} {\bibfnamefont {T.}~\bibnamefont
  {{Morinaga}}}, \bibinfo {author} {\bibfnamefont {H.}~\bibnamefont
  {{Nagakura}}}, \bibinfo {author} {\bibfnamefont {C.}~\bibnamefont {{Kato}}},\
  and\ \bibinfo {author} {\bibfnamefont {S.}~\bibnamefont {{Yamada}}},\
  }\bibfield  {title} {\bibinfo {title} {{Fast neutrino-flavor conversion in
  the preshock region of core-collapse supernovae}},\ }\href
  {https://doi.org/10.1103/PhysRevResearch.2.012046} {\bibfield  {journal}
  {\bibinfo  {journal} {Physical Review Research}\ }\textbf {\bibinfo {volume}
  {2}},\ \bibinfo {eid} {012046} (\bibinfo {year} {2020})},\ \Eprint
  {https://arxiv.org/abs/1909.13131} {arXiv:1909.13131 [astro-ph.HE]}
  \BibitemShut {NoStop}%
\bibitem [{\citenamefont {{Delfan Azari}}\ \emph {et~al.}(2020)\citenamefont
  {{Delfan Azari}}, \citenamefont {{Yamada}}, \citenamefont {{Morinaga}},
  \citenamefont {{Nagakura}}, \citenamefont {{Furusawa}}, \citenamefont
  {{Harada}}, \citenamefont {{Okawa}}, \citenamefont {{Iwakami}},\ and\
  \citenamefont {{Sumiyoshi}}}]{milad2020}%
  \BibitemOpen
  \bibfield  {author} {\bibinfo {author} {\bibfnamefont {M.}~\bibnamefont
  {{Delfan Azari}}}, \bibinfo {author} {\bibfnamefont {S.}~\bibnamefont
  {{Yamada}}}, \bibinfo {author} {\bibfnamefont {T.}~\bibnamefont
  {{Morinaga}}}, \bibinfo {author} {\bibfnamefont {H.}~\bibnamefont
  {{Nagakura}}}, \bibinfo {author} {\bibfnamefont {S.}~\bibnamefont
  {{Furusawa}}}, \bibinfo {author} {\bibfnamefont {A.}~\bibnamefont
  {{Harada}}}, \bibinfo {author} {\bibfnamefont {H.}~\bibnamefont {{Okawa}}},
  \bibinfo {author} {\bibfnamefont {W.}~\bibnamefont {{Iwakami}}},\ and\
  \bibinfo {author} {\bibfnamefont {K.}~\bibnamefont {{Sumiyoshi}}},\
  }\bibfield  {title} {\bibinfo {title} {{Fast collective neutrino oscillations
  inside the neutrino sphere in core-collapse supernovae}},\ }\href
  {https://doi.org/10.1103/PhysRevD.101.023018} {\bibfield  {journal} {\bibinfo
   {journal} {\prd}\ }\textbf {\bibinfo {volume} {101}},\ \bibinfo {eid}
  {023018} (\bibinfo {year} {2020})},\ \Eprint
  {https://arxiv.org/abs/1910.06176} {arXiv:1910.06176 [astro-ph.HE]}
  \BibitemShut {NoStop}%
\bibitem [{\citenamefont {Capozzi}\ \emph {et~al.}(2020)\citenamefont
  {Capozzi}, \citenamefont {Chakraborty}, \citenamefont {Chakraborty},\ and\
  \citenamefont {Sen}}]{capozzi2020}%
  \BibitemOpen
  \bibfield  {author} {\bibinfo {author} {\bibfnamefont {F.}~\bibnamefont
  {Capozzi}}, \bibinfo {author} {\bibfnamefont {M.}~\bibnamefont
  {Chakraborty}}, \bibinfo {author} {\bibfnamefont {S.}~\bibnamefont
  {Chakraborty}},\ and\ \bibinfo {author} {\bibfnamefont {M.}~\bibnamefont
  {Sen}},\ }\bibfield  {title} {\bibinfo {title} {Mu-tau neutrinos: Influencing
  fast flavor conversions in supernovae},\ }\href
  {https://doi.org/10.1103/PhysRevLett.125.251801} {\bibfield  {journal}
  {\bibinfo  {journal} {Phys. Rev. Lett.}\ }\textbf {\bibinfo {volume} {125}},\
  \bibinfo {pages} {251801} (\bibinfo {year} {2020})}\BibitemShut {NoStop}%
\bibitem [{\citenamefont {Abbar}\ \emph {et~al.}(2021)\citenamefont {Abbar},
  \citenamefont {Capozzi}, \citenamefont {Glas}, \citenamefont {Janka},\ and\
  \citenamefont {Tamborra}}]{abbar2021}%
  \BibitemOpen
  \bibfield  {author} {\bibinfo {author} {\bibfnamefont {S.}~\bibnamefont
  {Abbar}}, \bibinfo {author} {\bibfnamefont {F.}~\bibnamefont {Capozzi}},
  \bibinfo {author} {\bibfnamefont {R.}~\bibnamefont {Glas}}, \bibinfo {author}
  {\bibfnamefont {H.-T.}\ \bibnamefont {Janka}},\ and\ \bibinfo {author}
  {\bibfnamefont {I.}~\bibnamefont {Tamborra}},\ }\bibfield  {title} {\bibinfo
  {title} {On the characteristics of fast neutrino flavor instabilities in
  three-dimensional core-collapse supernova models},\ }\href
  {https://doi.org/10.1103/PhysRevD.103.063033} {\bibfield  {journal} {\bibinfo
   {journal} {Phys. Rev. D}\ }\textbf {\bibinfo {volume} {103}},\ \bibinfo
  {pages} {063033} (\bibinfo {year} {2021})}\BibitemShut {NoStop}%
\bibitem [{\citenamefont {{Capozzi}}\ \emph {et~al.}(2021)\citenamefont
  {{Capozzi}}, \citenamefont {{Abbar}}, \citenamefont {{Bollig}},\ and\
  \citenamefont {{Janka}}}]{capozzi2021}%
  \BibitemOpen
  \bibfield  {author} {\bibinfo {author} {\bibfnamefont {F.}~\bibnamefont
  {{Capozzi}}}, \bibinfo {author} {\bibfnamefont {S.}~\bibnamefont {{Abbar}}},
  \bibinfo {author} {\bibfnamefont {R.}~\bibnamefont {{Bollig}}},\ and\
  \bibinfo {author} {\bibfnamefont {H.~T.}\ \bibnamefont {{Janka}}},\
  }\bibfield  {title} {\bibinfo {title} {{Fast neutrino flavor conversions in
  one-dimensional core-collapse supernova models with and without muon
  creation}},\ }\href {https://doi.org/10.1103/PhysRevD.103.063013} {\bibfield
  {journal} {\bibinfo  {journal} {\prd}\ }\textbf {\bibinfo {volume} {103}},\
  \bibinfo {eid} {063013} (\bibinfo {year} {2021})},\ \Eprint
  {https://arxiv.org/abs/2012.08525} {arXiv:2012.08525 [astro-ph.HE]}
  \BibitemShut {NoStop}%
\bibitem [{\citenamefont {{Nagakura}}\ \emph {et~al.}(2021)\citenamefont
  {{Nagakura}}, \citenamefont {{Burrows}}, \citenamefont {{Johns}},\ and\
  \citenamefont {{Fuller}}}]{nagakura2021d}%
  \BibitemOpen
  \bibfield  {author} {\bibinfo {author} {\bibfnamefont {H.}~\bibnamefont
  {{Nagakura}}}, \bibinfo {author} {\bibfnamefont {A.}~\bibnamefont
  {{Burrows}}}, \bibinfo {author} {\bibfnamefont {L.}~\bibnamefont {{Johns}}},\
  and\ \bibinfo {author} {\bibfnamefont {G.~M.}\ \bibnamefont {{Fuller}}},\
  }\bibfield  {title} {\bibinfo {title} {{Where, when, and why: Occurrence of
  fast-pairwise collective neutrino oscillation in three-dimensional
  core-collapse supernova models}},\ }\href
  {https://doi.org/10.1103/PhysRevD.104.083025} {\bibfield  {journal} {\bibinfo
   {journal} {\prd}\ }\textbf {\bibinfo {volume} {104}},\ \bibinfo {eid}
  {083025} (\bibinfo {year} {2021})},\ \Eprint
  {https://arxiv.org/abs/2108.07281} {arXiv:2108.07281 [astro-ph.HE]}
  \BibitemShut {NoStop}%
\bibitem [{\citenamefont {{Harada}}\ and\ \citenamefont
  {{Nagakura}}(2022)}]{harada2022}%
  \BibitemOpen
  \bibfield  {author} {\bibinfo {author} {\bibfnamefont {A.}~\bibnamefont
  {{Harada}}}\ and\ \bibinfo {author} {\bibfnamefont {H.}~\bibnamefont
  {{Nagakura}}},\ }\bibfield  {title} {\bibinfo {title} {{Prospects of Fast
  Flavor Neutrino Conversion in Rotating Core-collapse Supernovae}},\ }\href
  {https://doi.org/10.3847/1538-4357/ac38a0} {\bibfield  {journal} {\bibinfo
  {journal} {\apj}\ }\textbf {\bibinfo {volume} {924}},\ \bibinfo {eid} {109}
  (\bibinfo {year} {2022})},\ \Eprint {https://arxiv.org/abs/2110.08291}
  {arXiv:2110.08291 [astro-ph.HE]} \BibitemShut {NoStop}%
\bibitem [{\citenamefont {{Richers}}(2022)}]{sherwood2022}%
  \BibitemOpen
  \bibfield  {author} {\bibinfo {author} {\bibfnamefont {S.}~\bibnamefont
  {{Richers}}},\ }\bibfield  {title} {\bibinfo {title} {{Evaluating approximate
  flavor instability metrics in neutron star mergers}},\ }\href
  {https://doi.org/10.1103/PhysRevD.106.083005} {\bibfield  {journal} {\bibinfo
   {journal} {\prd}\ }\textbf {\bibinfo {volume} {106}},\ \bibinfo {eid}
  {083005} (\bibinfo {year} {2022})},\ \Eprint
  {https://arxiv.org/abs/2206.08444} {arXiv:2206.08444 [astro-ph.HE]}
  \BibitemShut {NoStop}%
\bibitem [{\citenamefont {{Akaho}}\ \emph {et~al.}(2022)\citenamefont
  {{Akaho}}, \citenamefont {{Harada}}, \citenamefont {{Nagakura}},
  \citenamefont {{Iwakami}}, \citenamefont {{Okawa}}, \citenamefont
  {{Furusawa}}, \citenamefont {{Matsufuru}}, \citenamefont {{Sumiyoshi}},\ and\
  \citenamefont {{Yamada}}}]{akaho2022}%
  \BibitemOpen
  \bibfield  {author} {\bibinfo {author} {\bibfnamefont {R.}~\bibnamefont
  {{Akaho}}}, \bibinfo {author} {\bibfnamefont {A.}~\bibnamefont {{Harada}}},
  \bibinfo {author} {\bibfnamefont {H.}~\bibnamefont {{Nagakura}}}, \bibinfo
  {author} {\bibfnamefont {W.}~\bibnamefont {{Iwakami}}}, \bibinfo {author}
  {\bibfnamefont {H.}~\bibnamefont {{Okawa}}}, \bibinfo {author} {\bibfnamefont
  {S.}~\bibnamefont {{Furusawa}}}, \bibinfo {author} {\bibfnamefont
  {H.}~\bibnamefont {{Matsufuru}}}, \bibinfo {author} {\bibfnamefont
  {K.}~\bibnamefont {{Sumiyoshi}}},\ and\ \bibinfo {author} {\bibfnamefont
  {S.}~\bibnamefont {{Yamada}}},\ }\bibfield  {title} {\bibinfo {title}
  {{Protoneutron Star Convection Simulated with a New General Relativistic
  Boltzmann Neutrino Radiation-Hydrodynamics Code}},\ }\href@noop {} {\bibfield
   {journal} {\bibinfo  {journal} {arXiv e-prints}\ ,\ \bibinfo {eid}
  {arXiv:2206.01673}} (\bibinfo {year} {2022})},\ \Eprint
  {https://arxiv.org/abs/2206.01673} {arXiv:2206.01673 [astro-ph.HE]}
  \BibitemShut {NoStop}%
\bibitem [{\citenamefont {George}\ \emph {et~al.}(2020)\citenamefont {George},
  \citenamefont {Wu}, \citenamefont {Tamborra}, \citenamefont
  {Ardevol-Pulpillo},\ and\ \citenamefont {Janka}}]{george2020}%
  \BibitemOpen
  \bibfield  {author} {\bibinfo {author} {\bibfnamefont {M.}~\bibnamefont
  {George}}, \bibinfo {author} {\bibfnamefont {M.-R.}\ \bibnamefont {Wu}},
  \bibinfo {author} {\bibfnamefont {I.}~\bibnamefont {Tamborra}}, \bibinfo
  {author} {\bibfnamefont {R.}~\bibnamefont {Ardevol-Pulpillo}},\ and\ \bibinfo
  {author} {\bibfnamefont {H.-T.}\ \bibnamefont {Janka}},\ }\bibfield  {title}
  {\bibinfo {title} {Fast neutrino flavor conversion, ejecta properties, and
  nucleosynthesis in newly-formed hypermassive remnants of neutron-star
  mergers},\ }\href {https://doi.org/10.1103/PhysRevD.102.103015} {\bibfield
  {journal} {\bibinfo  {journal} {Phys. Rev. D}\ }\textbf {\bibinfo {volume}
  {102}},\ \bibinfo {pages} {103015} (\bibinfo {year} {2020})}\BibitemShut
  {NoStop}%
\bibitem [{\citenamefont {{Li}}\ and\ \citenamefont {{Siegel}}(2021)}]{Li2021}%
  \BibitemOpen
  \bibfield  {author} {\bibinfo {author} {\bibfnamefont {X.}~\bibnamefont
  {{Li}}}\ and\ \bibinfo {author} {\bibfnamefont {D.~M.}\ \bibnamefont
  {{Siegel}}},\ }\bibfield  {title} {\bibinfo {title} {{Neutrino Fast Flavor
  Conversions in Neutron-Star Postmerger Accretion Disks}},\ }\href
  {https://doi.org/10.1103/PhysRevLett.126.251101} {\bibfield  {journal}
  {\bibinfo  {journal} {\prl}\ }\textbf {\bibinfo {volume} {126}},\ \bibinfo
  {eid} {251101} (\bibinfo {year} {2021})},\ \Eprint
  {https://arxiv.org/abs/2103.02616} {arXiv:2103.02616 [astro-ph.HE]}
  \BibitemShut {NoStop}%
\bibitem [{\citenamefont {{Just}}\ \emph {et~al.}(2022)\citenamefont {{Just}},
  \citenamefont {{Abbar}}, \citenamefont {{Wu}}, \citenamefont {{Tamborra}},
  \citenamefont {{Janka}},\ and\ \citenamefont {{Capozzi}}}]{Just2022}%
  \BibitemOpen
  \bibfield  {author} {\bibinfo {author} {\bibfnamefont {O.}~\bibnamefont
  {{Just}}}, \bibinfo {author} {\bibfnamefont {S.}~\bibnamefont {{Abbar}}},
  \bibinfo {author} {\bibfnamefont {M.-R.}\ \bibnamefont {{Wu}}}, \bibinfo
  {author} {\bibfnamefont {I.}~\bibnamefont {{Tamborra}}}, \bibinfo {author}
  {\bibfnamefont {H.-T.}\ \bibnamefont {{Janka}}},\ and\ \bibinfo {author}
  {\bibfnamefont {F.}~\bibnamefont {{Capozzi}}},\ }\bibfield  {title} {\bibinfo
  {title} {{Fast neutrino conversion in hydrodynamic simulations of
  neutrino-cooled accretion disks}},\ }\href
  {https://doi.org/10.1103/PhysRevD.105.083024} {\bibfield  {journal} {\bibinfo
   {journal} {\prd}\ }\textbf {\bibinfo {volume} {105}},\ \bibinfo {eid}
  {083024} (\bibinfo {year} {2022})},\ \Eprint
  {https://arxiv.org/abs/2203.16559} {arXiv:2203.16559 [astro-ph.HE]}
  \BibitemShut {NoStop}%
\bibitem [{\citenamefont {{Abbar}}\ and\ \citenamefont
  {{Volpe}}(2019)}]{abbar2019}%
  \BibitemOpen
  \bibfield  {author} {\bibinfo {author} {\bibfnamefont {S.}~\bibnamefont
  {{Abbar}}}\ and\ \bibinfo {author} {\bibfnamefont {M.~C.}\ \bibnamefont
  {{Volpe}}},\ }\bibfield  {title} {\bibinfo {title} {{On fast neutrino flavor
  conversion modes in the nonlinear regime}},\ }\href
  {https://doi.org/10.1016/j.physletb.2019.02.002} {\bibfield  {journal}
  {\bibinfo  {journal} {Physics Letters B}\ }\textbf {\bibinfo {volume}
  {790}},\ \bibinfo {pages} {545} (\bibinfo {year} {2019})},\ \Eprint
  {https://arxiv.org/abs/1811.04215} {arXiv:1811.04215 [astro-ph.HE]}
  \BibitemShut {NoStop}%
\bibitem [{\citenamefont {{Martin}}\ \emph {et~al.}(2020)\citenamefont
  {{Martin}}, \citenamefont {{Yi}},\ and\ \citenamefont {{Duan}}}]{martin2020}%
  \BibitemOpen
  \bibfield  {author} {\bibinfo {author} {\bibfnamefont {J.~D.}\ \bibnamefont
  {{Martin}}}, \bibinfo {author} {\bibfnamefont {C.}~\bibnamefont {{Yi}}},\
  and\ \bibinfo {author} {\bibfnamefont {H.}~\bibnamefont {{Duan}}},\
  }\bibfield  {title} {\bibinfo {title} {{Dynamic fast flavor oscillation waves
  in dense neutrino gases}},\ }\href
  {https://doi.org/10.1016/j.physletb.2019.135088} {\bibfield  {journal}
  {\bibinfo  {journal} {Physics Letters B}\ }\textbf {\bibinfo {volume}
  {800}},\ \bibinfo {eid} {135088} (\bibinfo {year} {2020})},\ \Eprint
  {https://arxiv.org/abs/1909.05225} {arXiv:1909.05225 [hep-ph]} \BibitemShut
  {NoStop}%
\bibitem [{\citenamefont {{Johns}}\ \emph
  {et~al.}(2020{\natexlab{a}})\citenamefont {{Johns}}, \citenamefont
  {{Nagakura}}, \citenamefont {{Fuller}},\ and\ \citenamefont
  {{Burrows}}}]{johns2020b}%
  \BibitemOpen
  \bibfield  {author} {\bibinfo {author} {\bibfnamefont {L.}~\bibnamefont
  {{Johns}}}, \bibinfo {author} {\bibfnamefont {H.}~\bibnamefont {{Nagakura}}},
  \bibinfo {author} {\bibfnamefont {G.~M.}\ \bibnamefont {{Fuller}}},\ and\
  \bibinfo {author} {\bibfnamefont {A.}~\bibnamefont {{Burrows}}},\ }\bibfield
  {title} {\bibinfo {title} {{Fast oscillations, collisionless relaxation, and
  spurious evolution of supernova neutrino flavor}},\ }\href
  {https://doi.org/10.1103/PhysRevD.102.103017} {\bibfield  {journal} {\bibinfo
   {journal} {\prd}\ }\textbf {\bibinfo {volume} {102}},\ \bibinfo {eid}
  {103017} (\bibinfo {year} {2020}{\natexlab{a}})},\ \Eprint
  {https://arxiv.org/abs/2009.09024} {arXiv:2009.09024 [hep-ph]} \BibitemShut
  {NoStop}%
\bibitem [{\citenamefont {Bhattacharyya}\ and\ \citenamefont
  {Dasgupta}(2020)}]{bhattacharyya2020}%
  \BibitemOpen
  \bibfield  {author} {\bibinfo {author} {\bibfnamefont {S.}~\bibnamefont
  {Bhattacharyya}}\ and\ \bibinfo {author} {\bibfnamefont {B.}~\bibnamefont
  {Dasgupta}},\ }\bibfield  {title} {\bibinfo {title} {Late-time behavior of
  fast neutrino oscillations},\ }\href
  {https://doi.org/10.1103/PhysRevD.102.063018} {\bibfield  {journal} {\bibinfo
   {journal} {Phys. Rev. D}\ }\textbf {\bibinfo {volume} {102}},\ \bibinfo
  {pages} {063018} (\bibinfo {year} {2020})}\BibitemShut {NoStop}%
\bibitem [{\citenamefont {{Bhattacharyya}}\ and\ \citenamefont
  {{Dasgupta}}(2021)}]{bhattacharayya2021}%
  \BibitemOpen
  \bibfield  {author} {\bibinfo {author} {\bibfnamefont {S.}~\bibnamefont
  {{Bhattacharyya}}}\ and\ \bibinfo {author} {\bibfnamefont {B.}~\bibnamefont
  {{Dasgupta}}},\ }\bibfield  {title} {\bibinfo {title} {{Fast Flavor
  Depolarization of Supernova Neutrinos}},\ }\href
  {https://doi.org/10.1103/PhysRevLett.126.061302} {\bibfield  {journal}
  {\bibinfo  {journal} {\prl}\ }\textbf {\bibinfo {volume} {126}},\ \bibinfo
  {eid} {061302} (\bibinfo {year} {2021})},\ \Eprint
  {https://arxiv.org/abs/2009.03337} {arXiv:2009.03337 [hep-ph]} \BibitemShut
  {NoStop}%
\bibitem [{\citenamefont {Richers}\ \emph {et~al.}(2021)\citenamefont
  {Richers}, \citenamefont {Willcox}, \citenamefont {Ford},\ and\ \citenamefont
  {Myers}}]{sherwood2021}%
  \BibitemOpen
  \bibfield  {author} {\bibinfo {author} {\bibfnamefont {S.}~\bibnamefont
  {Richers}}, \bibinfo {author} {\bibfnamefont {D.~E.}\ \bibnamefont
  {Willcox}}, \bibinfo {author} {\bibfnamefont {N.~M.}\ \bibnamefont {Ford}},\
  and\ \bibinfo {author} {\bibfnamefont {A.}~\bibnamefont {Myers}},\ }\bibfield
   {title} {\bibinfo {title} {Particle-in-cell simulation of the neutrino fast
  flavor instability},\ }\href {https://doi.org/10.1103/PhysRevD.103.083013}
  {\bibfield  {journal} {\bibinfo  {journal} {Phys. Rev. D}\ }\textbf {\bibinfo
  {volume} {103}},\ \bibinfo {pages} {083013} (\bibinfo {year}
  {2021})}\BibitemShut {NoStop}%
\bibitem [{\citenamefont {{Richers}}\ \emph {et~al.}(2021)\citenamefont
  {{Richers}}, \citenamefont {{Willcox}},\ and\ \citenamefont
  {{Ford}}}]{sherwood2021b}%
  \BibitemOpen
  \bibfield  {author} {\bibinfo {author} {\bibfnamefont {S.}~\bibnamefont
  {{Richers}}}, \bibinfo {author} {\bibfnamefont {D.}~\bibnamefont
  {{Willcox}}},\ and\ \bibinfo {author} {\bibfnamefont {N.}~\bibnamefont
  {{Ford}}},\ }\bibfield  {title} {\bibinfo {title} {{Neutrino fast flavor
  instability in three dimensions}},\ }\href
  {https://doi.org/10.1103/PhysRevD.104.103023} {\bibfield  {journal} {\bibinfo
   {journal} {\prd}\ }\textbf {\bibinfo {volume} {104}},\ \bibinfo {eid}
  {103023} (\bibinfo {year} {2021})},\ \Eprint
  {https://arxiv.org/abs/2109.08631} {arXiv:2109.08631 [astro-ph.HE]}
  \BibitemShut {NoStop}%
\bibitem [{\citenamefont {{Wu}}\ \emph {et~al.}(2021)\citenamefont {{Wu}},
  \citenamefont {{George}}, \citenamefont {{Lin}},\ and\ \citenamefont
  {{Xiong}}}]{wu2021}%
  \BibitemOpen
  \bibfield  {author} {\bibinfo {author} {\bibfnamefont {M.-R.}\ \bibnamefont
  {{Wu}}}, \bibinfo {author} {\bibfnamefont {M.}~\bibnamefont {{George}}},
  \bibinfo {author} {\bibfnamefont {C.-Y.}\ \bibnamefont {{Lin}}},\ and\
  \bibinfo {author} {\bibfnamefont {Z.}~\bibnamefont {{Xiong}}},\ }\bibfield
  {title} {\bibinfo {title} {{Collective fast neutrino flavor conversions in a
  1D box: Initial conditions and long-term evolution}},\ }\href
  {https://doi.org/10.1103/PhysRevD.104.103003} {\bibfield  {journal} {\bibinfo
   {journal} {\prd}\ }\textbf {\bibinfo {volume} {104}},\ \bibinfo {eid}
  {103003} (\bibinfo {year} {2021})},\ \Eprint
  {https://arxiv.org/abs/2108.09886} {arXiv:2108.09886 [hep-ph]} \BibitemShut
  {NoStop}%
\bibitem [{\citenamefont {{Richers}}\ \emph {et~al.}(2019)\citenamefont
  {{Richers}}, \citenamefont {{McLaughlin}}, \citenamefont {{Kneller}},\ and\
  \citenamefont {{Vlasenko}}}]{sherwood2019}%
  \BibitemOpen
  \bibfield  {author} {\bibinfo {author} {\bibfnamefont {S.~A.}\ \bibnamefont
  {{Richers}}}, \bibinfo {author} {\bibfnamefont {G.~C.}\ \bibnamefont
  {{McLaughlin}}}, \bibinfo {author} {\bibfnamefont {J.~P.}\ \bibnamefont
  {{Kneller}}},\ and\ \bibinfo {author} {\bibfnamefont {A.}~\bibnamefont
  {{Vlasenko}}},\ }\bibfield  {title} {\bibinfo {title} {{Neutrino quantum
  kinetics in compact objects}},\ }\href
  {https://doi.org/10.1103/PhysRevD.99.123014} {\bibfield  {journal} {\bibinfo
  {journal} {\prd}\ }\textbf {\bibinfo {volume} {99}},\ \bibinfo {eid} {123014}
  (\bibinfo {year} {2019})},\ \Eprint {https://arxiv.org/abs/1903.00022}
  {arXiv:1903.00022 [astro-ph.HE]} \BibitemShut {NoStop}%
\bibitem [{\citenamefont {{Shalgar}}\ and\ \citenamefont
  {{Tamborra}}(2021{\natexlab{a}})}]{shalgar2021}%
  \BibitemOpen
  \bibfield  {author} {\bibinfo {author} {\bibfnamefont {S.}~\bibnamefont
  {{Shalgar}}}\ and\ \bibinfo {author} {\bibfnamefont {I.}~\bibnamefont
  {{Tamborra}}},\ }\bibfield  {title} {\bibinfo {title} {{Dispelling a myth on
  dense neutrino media: fast pairwise conversions depend on energy}},\ }\href
  {https://doi.org/10.1088/1475-7516/2021/01/014} {\bibfield  {journal}
  {\bibinfo  {journal} {\jcap}\ }\textbf {\bibinfo {volume} {2021}},\ \bibinfo
  {eid} {014} (\bibinfo {year} {2021}{\natexlab{a}})},\ \Eprint
  {https://arxiv.org/abs/2007.07926} {arXiv:2007.07926 [astro-ph.HE]}
  \BibitemShut {NoStop}%
\bibitem [{\citenamefont {{Martin}}\ \emph {et~al.}(2021)\citenamefont
  {{Martin}}, \citenamefont {{Carlson}}, \citenamefont {{Cirigliano}},\ and\
  \citenamefont {{Duan}}}]{martin2021}%
  \BibitemOpen
  \bibfield  {author} {\bibinfo {author} {\bibfnamefont {J.~D.}\ \bibnamefont
  {{Martin}}}, \bibinfo {author} {\bibfnamefont {J.}~\bibnamefont {{Carlson}}},
  \bibinfo {author} {\bibfnamefont {V.}~\bibnamefont {{Cirigliano}}},\ and\
  \bibinfo {author} {\bibfnamefont {H.}~\bibnamefont {{Duan}}},\ }\bibfield
  {title} {\bibinfo {title} {{Fast flavor oscillations in dense neutrino media
  with collisions}},\ }\href {https://doi.org/10.1103/PhysRevD.103.063001}
  {\bibfield  {journal} {\bibinfo  {journal} {\prd}\ }\textbf {\bibinfo
  {volume} {103}},\ \bibinfo {eid} {063001} (\bibinfo {year} {2021})},\ \Eprint
  {https://arxiv.org/abs/2101.01278} {arXiv:2101.01278 [hep-ph]} \BibitemShut
  {NoStop}%
\bibitem [{\citenamefont {{Zaizen}}\ and\ \citenamefont
  {{Morinaga}}(2021)}]{zaizen2021}%
  \BibitemOpen
  \bibfield  {author} {\bibinfo {author} {\bibfnamefont {M.}~\bibnamefont
  {{Zaizen}}}\ and\ \bibinfo {author} {\bibfnamefont {T.}~\bibnamefont
  {{Morinaga}}},\ }\bibfield  {title} {\bibinfo {title} {{Nonlinear evolution
  of fast neutrino flavor conversion in the preshock region of core-collapse
  supernovae}},\ }\href {https://doi.org/10.1103/PhysRevD.104.083035}
  {\bibfield  {journal} {\bibinfo  {journal} {\prd}\ }\textbf {\bibinfo
  {volume} {104}},\ \bibinfo {eid} {083035} (\bibinfo {year} {2021})},\ \Eprint
  {https://arxiv.org/abs/2104.10532} {arXiv:2104.10532 [hep-ph]} \BibitemShut
  {NoStop}%
\bibitem [{\citenamefont {{Bhattacharyya}}\ and\ \citenamefont
  {{Dasgupta}}(2022)}]{bhttacharyya2022}%
  \BibitemOpen
  \bibfield  {author} {\bibinfo {author} {\bibfnamefont {S.}~\bibnamefont
  {{Bhattacharyya}}}\ and\ \bibinfo {author} {\bibfnamefont {B.}~\bibnamefont
  {{Dasgupta}}},\ }\bibfield  {title} {\bibinfo {title} {{Elaborating the
  Ultimate Fate of Fast Collective Neutrino Flavor Oscillations}},\ }\href@noop
  {} {\bibfield  {journal} {\bibinfo  {journal} {arXiv e-prints}\ ,\ \bibinfo
  {eid} {arXiv:2205.05129}} (\bibinfo {year} {2022})},\ \Eprint
  {https://arxiv.org/abs/2205.05129} {arXiv:2205.05129 [hep-ph]} \BibitemShut
  {NoStop}%
\bibitem [{\citenamefont {{Abbar}}\ and\ \citenamefont
  {{Capozzi}}(2022)}]{Abbar2022}%
  \BibitemOpen
  \bibfield  {author} {\bibinfo {author} {\bibfnamefont {S.}~\bibnamefont
  {{Abbar}}}\ and\ \bibinfo {author} {\bibfnamefont {F.}~\bibnamefont
  {{Capozzi}}},\ }\bibfield  {title} {\bibinfo {title} {{Suppression of fast
  neutrino flavor conversions occurring at large distances in core-collapse
  supernovae}},\ }\href {https://doi.org/10.1088/1475-7516/2022/03/051}
  {\bibfield  {journal} {\bibinfo  {journal} {\jcap}\ }\textbf {\bibinfo
  {volume} {2022}},\ \bibinfo {eid} {051} (\bibinfo {year} {2022})},\ \Eprint
  {https://arxiv.org/abs/2111.14880} {arXiv:2111.14880 [astro-ph.HE]}
  \BibitemShut {NoStop}%
\bibitem [{\citenamefont {{Nagakura}}\ and\ \citenamefont
  {{Zaizen}}(2022)}]{nagakura2022}%
  \BibitemOpen
  \bibfield  {author} {\bibinfo {author} {\bibfnamefont {H.}~\bibnamefont
  {{Nagakura}}}\ and\ \bibinfo {author} {\bibfnamefont {M.}~\bibnamefont
  {{Zaizen}}},\ }\bibfield  {title} {\bibinfo {title} {{Time-dependent,
  quasi-steady, and global features of fast neutrino-flavor conversion}},\
  }\href@noop {} {\bibfield  {journal} {\bibinfo  {journal} {arXiv e-prints}\
  ,\ \bibinfo {eid} {arXiv:2206.04097}} (\bibinfo {year} {2022})},\ \Eprint
  {https://arxiv.org/abs/2206.04097} {arXiv:2206.04097 [astro-ph.HE]}
  \BibitemShut {NoStop}%
\bibitem [{\citenamefont {{Shalgar}}\ and\ \citenamefont
  {{Tamborra}}(2022{\natexlab{a}})}]{shalgar2022}%
  \BibitemOpen
  \bibfield  {author} {\bibinfo {author} {\bibfnamefont {S.}~\bibnamefont
  {{Shalgar}}}\ and\ \bibinfo {author} {\bibfnamefont {I.}~\bibnamefont
  {{Tamborra}}},\ }\bibfield  {title} {\bibinfo {title} {{Supernova Neutrino
  Decoupling Is Altered by Flavor Conversion}},\ }\href@noop {} {\bibfield
  {journal} {\bibinfo  {journal} {arXiv e-prints}\ ,\ \bibinfo {eid}
  {arXiv:2206.00676}} (\bibinfo {year} {2022}{\natexlab{a}})},\ \Eprint
  {https://arxiv.org/abs/2206.00676} {arXiv:2206.00676 [astro-ph.HE]}
  \BibitemShut {NoStop}%
\bibitem [{\citenamefont {{Shalgar}}\ and\ \citenamefont
  {{Tamborra}}(2022{\natexlab{b}})}]{shalgar2022b}%
  \BibitemOpen
  \bibfield  {author} {\bibinfo {author} {\bibfnamefont {S.}~\bibnamefont
  {{Shalgar}}}\ and\ \bibinfo {author} {\bibfnamefont {I.}~\bibnamefont
  {{Tamborra}}},\ }\bibfield  {title} {\bibinfo {title} {{Neutrino Flavor
  Conversion, Advection, and Collisions: The Full Solution}},\ }\href@noop {}
  {\bibfield  {journal} {\bibinfo  {journal} {arXiv e-prints}\ ,\ \bibinfo
  {eid} {arXiv:2207.04058}} (\bibinfo {year} {2022}{\natexlab{b}})},\ \Eprint
  {https://arxiv.org/abs/2207.04058} {arXiv:2207.04058 [astro-ph.HE]}
  \BibitemShut {NoStop}%
\bibitem [{\citenamefont {{Shalgar}}\ and\ \citenamefont
  {{Tamborra}}(2021{\natexlab{b}})}]{shalgar2021b}%
  \BibitemOpen
  \bibfield  {author} {\bibinfo {author} {\bibfnamefont {S.}~\bibnamefont
  {{Shalgar}}}\ and\ \bibinfo {author} {\bibfnamefont {I.}~\bibnamefont
  {{Tamborra}}},\ }\bibfield  {title} {\bibinfo {title} {{Change of direction
  in pairwise neutrino conversion physics: The effect of collisions}},\ }\href
  {https://doi.org/10.1103/PhysRevD.103.063002} {\bibfield  {journal} {\bibinfo
   {journal} {\prd}\ }\textbf {\bibinfo {volume} {103}},\ \bibinfo {eid}
  {063002} (\bibinfo {year} {2021}{\natexlab{b}})},\ \Eprint
  {https://arxiv.org/abs/2011.00004} {arXiv:2011.00004 [astro-ph.HE]}
  \BibitemShut {NoStop}%
\bibitem [{\citenamefont {{Kato}}\ \emph {et~al.}(2021)\citenamefont {{Kato}},
  \citenamefont {{Nagakura}},\ and\ \citenamefont {{Morinaga}}}]{kato2021}%
  \BibitemOpen
  \bibfield  {author} {\bibinfo {author} {\bibfnamefont {C.}~\bibnamefont
  {{Kato}}}, \bibinfo {author} {\bibfnamefont {H.}~\bibnamefont {{Nagakura}}},\
  and\ \bibinfo {author} {\bibfnamefont {T.}~\bibnamefont {{Morinaga}}},\
  }\bibfield  {title} {\bibinfo {title} {{Neutrino Transport with the Monte
  Carlo Method. II. Quantum Kinetic Equations}},\ }\href
  {https://doi.org/10.3847/1538-4365/ac2aa4} {\bibfield  {journal} {\bibinfo
  {journal} {\apjs}\ }\textbf {\bibinfo {volume} {257}},\ \bibinfo {eid} {55}
  (\bibinfo {year} {2021})},\ \Eprint {https://arxiv.org/abs/2108.06356}
  {arXiv:2108.06356 [astro-ph.HE]} \BibitemShut {NoStop}%
\bibitem [{\citenamefont {{Sasaki}}\ and\ \citenamefont
  {{Takiwaki}}(2022)}]{sasaki2021}%
  \BibitemOpen
  \bibfield  {author} {\bibinfo {author} {\bibfnamefont {H.}~\bibnamefont
  {{Sasaki}}}\ and\ \bibinfo {author} {\bibfnamefont {T.}~\bibnamefont
  {{Takiwaki}}},\ }\bibfield  {title} {\bibinfo {title} {{A detailed analysis
  of the dynamics of fast neutrino flavor conversions with scattering
  effects}},\ }\href {https://doi.org/10.1093/ptep/ptac082} {\bibfield
  {journal} {\bibinfo  {journal} {Progress of Theoretical and Experimental
  Physics}\ }\textbf {\bibinfo {volume} {2022}},\ \bibinfo {eid} {073E01}
  (\bibinfo {year} {2022})},\ \Eprint {https://arxiv.org/abs/2109.14011}
  {arXiv:2109.14011 [hep-ph]} \BibitemShut {NoStop}%
\bibitem [{\citenamefont {{Sigl}}(2022)}]{sigl2022}%
  \BibitemOpen
  \bibfield  {author} {\bibinfo {author} {\bibfnamefont {G.}~\bibnamefont
  {{Sigl}}},\ }\bibfield  {title} {\bibinfo {title} {{Simulations of fast
  neutrino flavor conversions with interactions in inhomogeneous media}},\
  }\href {https://doi.org/10.1103/PhysRevD.105.043005} {\bibfield  {journal}
  {\bibinfo  {journal} {\prd}\ }\textbf {\bibinfo {volume} {105}},\ \bibinfo
  {eid} {043005} (\bibinfo {year} {2022})},\ \Eprint
  {https://arxiv.org/abs/2109.00091} {arXiv:2109.00091 [hep-ph]} \BibitemShut
  {NoStop}%
\bibitem [{\citenamefont {{Hansen}}\ \emph {et~al.}(2022)\citenamefont
  {{Hansen}}, \citenamefont {{Shalgar}},\ and\ \citenamefont
  {{Tamborra}}}]{hansen2022}%
  \BibitemOpen
  \bibfield  {author} {\bibinfo {author} {\bibfnamefont {R.~S.~L.}\
  \bibnamefont {{Hansen}}}, \bibinfo {author} {\bibfnamefont {S.}~\bibnamefont
  {{Shalgar}}},\ and\ \bibinfo {author} {\bibfnamefont {I.}~\bibnamefont
  {{Tamborra}}},\ }\bibfield  {title} {\bibinfo {title} {{Enhancement or
  damping of fast neutrino flavor conversions due to collisions}},\ }\href
  {https://doi.org/10.1103/PhysRevD.105.123003} {\bibfield  {journal} {\bibinfo
   {journal} {\prd}\ }\textbf {\bibinfo {volume} {105}},\ \bibinfo {eid}
  {123003} (\bibinfo {year} {2022})},\ \Eprint
  {https://arxiv.org/abs/2204.11873} {arXiv:2204.11873 [astro-ph.HE]}
  \BibitemShut {NoStop}%
\bibitem [{\citenamefont {{Johns}}\ and\ \citenamefont
  {{Nagakura}}(2022)}]{lucas2022}%
  \BibitemOpen
  \bibfield  {author} {\bibinfo {author} {\bibfnamefont {L.}~\bibnamefont
  {{Johns}}}\ and\ \bibinfo {author} {\bibfnamefont {H.}~\bibnamefont
  {{Nagakura}}},\ }\bibfield  {title} {\bibinfo {title} {{Self-consistency in
  models of neutrino scattering and fast flavor conversion}},\ }\href
  {https://doi.org/10.1103/PhysRevD.106.043031} {\bibfield  {journal} {\bibinfo
   {journal} {\prd}\ }\textbf {\bibinfo {volume} {106}},\ \bibinfo {eid}
  {043031} (\bibinfo {year} {2022})},\ \Eprint
  {https://arxiv.org/abs/2206.09225} {arXiv:2206.09225 [hep-ph]} \BibitemShut
  {NoStop}%
\bibitem [{\citenamefont {{Johns}}(2021)}]{johns2021}%
  \BibitemOpen
  \bibfield  {author} {\bibinfo {author} {\bibfnamefont {L.}~\bibnamefont
  {{Johns}}},\ }\bibfield  {title} {\bibinfo {title} {{Collisional flavor
  instabilities of supernova neutrinos}},\ }\href@noop {} {\bibfield  {journal}
  {\bibinfo  {journal} {arXiv e-prints}\ ,\ \bibinfo {eid} {arXiv:2104.11369}}
  (\bibinfo {year} {2021})},\ \Eprint {https://arxiv.org/abs/2104.11369}
  {arXiv:2104.11369 [hep-ph]} \BibitemShut {NoStop}%
\bibitem [{\citenamefont {{Volpe}}(2015)}]{volpe2015}%
  \BibitemOpen
  \bibfield  {author} {\bibinfo {author} {\bibfnamefont {C.}~\bibnamefont
  {{Volpe}}},\ }\bibfield  {title} {\bibinfo {title} {{Neutrino quantum kinetic
  equations}},\ }\href {https://doi.org/10.1142/S0218301315410098} {\bibfield
  {journal} {\bibinfo  {journal} {International Journal of Modern Physics E}\
  }\textbf {\bibinfo {volume} {24}},\ \bibinfo {eid} {1541009} (\bibinfo {year}
  {2015})},\ \Eprint {https://arxiv.org/abs/1506.06222} {arXiv:1506.06222
  [astro-ph.SR]} \BibitemShut {NoStop}%
\bibitem [{\citenamefont {{Kato}}\ \emph {et~al.}(2020)\citenamefont {{Kato}},
  \citenamefont {{Nagakura}}, \citenamefont {{Hori}},\ and\ \citenamefont
  {{Yamada}}}]{kato2020}%
  \BibitemOpen
  \bibfield  {author} {\bibinfo {author} {\bibfnamefont {C.}~\bibnamefont
  {{Kato}}}, \bibinfo {author} {\bibfnamefont {H.}~\bibnamefont {{Nagakura}}},
  \bibinfo {author} {\bibfnamefont {Y.}~\bibnamefont {{Hori}}},\ and\ \bibinfo
  {author} {\bibfnamefont {S.}~\bibnamefont {{Yamada}}},\ }\bibfield  {title}
  {\bibinfo {title} {{Neutrino Transport with Monte Carlo Method. I. Toward
  Fully Consistent Implementation of Nucleon Recoils in Core-collapse Supernova
  Simulations}},\ }\href {https://doi.org/10.3847/1538-4357/ab97b2} {\bibfield
  {journal} {\bibinfo  {journal} {\apj}\ }\textbf {\bibinfo {volume} {897}},\
  \bibinfo {eid} {43} (\bibinfo {year} {2020})},\ \Eprint
  {https://arxiv.org/abs/2001.11148} {arXiv:2001.11148 [astro-ph.HE]}
  \BibitemShut {NoStop}%
\bibitem [{\citenamefont {{Padilla-Gay}}\ \emph {et~al.}(2022)\citenamefont
  {{Padilla-Gay}}, \citenamefont {{Tamborra}},\ and\ \citenamefont
  {{Raffelt}}}]{Ian2022}%
  \BibitemOpen
  \bibfield  {author} {\bibinfo {author} {\bibfnamefont {I.}~\bibnamefont
  {{Padilla-Gay}}}, \bibinfo {author} {\bibfnamefont {I.}~\bibnamefont
  {{Tamborra}}},\ and\ \bibinfo {author} {\bibfnamefont {G.~G.}\ \bibnamefont
  {{Raffelt}}},\ }\bibfield  {title} {\bibinfo {title} {{Neutrino Flavor
  Pendulum Reloaded: The Case of Fast Pairwise Conversion}},\ }\href
  {https://doi.org/10.1103/PhysRevLett.128.121102} {\bibfield  {journal}
  {\bibinfo  {journal} {\prl}\ }\textbf {\bibinfo {volume} {128}},\ \bibinfo
  {eid} {121102} (\bibinfo {year} {2022})},\ \Eprint
  {https://arxiv.org/abs/2109.14627} {arXiv:2109.14627 [astro-ph.HE]}
  \BibitemShut {NoStop}%
\bibitem [{\citenamefont {{Johns}}\ \emph
  {et~al.}(2020{\natexlab{b}})\citenamefont {{Johns}}, \citenamefont
  {{Nagakura}}, \citenamefont {{Fuller}},\ and\ \citenamefont
  {{Burrows}}}]{johns2020}%
  \BibitemOpen
  \bibfield  {author} {\bibinfo {author} {\bibfnamefont {L.}~\bibnamefont
  {{Johns}}}, \bibinfo {author} {\bibfnamefont {H.}~\bibnamefont {{Nagakura}}},
  \bibinfo {author} {\bibfnamefont {G.~M.}\ \bibnamefont {{Fuller}}},\ and\
  \bibinfo {author} {\bibfnamefont {A.}~\bibnamefont {{Burrows}}},\ }\bibfield
  {title} {\bibinfo {title} {{Neutrino oscillations in supernovae: Angular
  moments and fast instabilities}},\ }\href
  {https://doi.org/10.1103/PhysRevD.101.043009} {\bibfield  {journal} {\bibinfo
   {journal} {\prd}\ }\textbf {\bibinfo {volume} {101}},\ \bibinfo {eid}
  {043009} (\bibinfo {year} {2020}{\natexlab{b}})},\ \Eprint
  {https://arxiv.org/abs/1910.05682} {arXiv:1910.05682 [hep-ph]} \BibitemShut
  {NoStop}%
\end{thebibliography}%

\end{document}